\documentclass[twocolumn,aps,showpacs,nofootinbib,prd]{revtex4-1}
\usepackage{graphicx}
\usepackage{amsbsy}
\usepackage{color}
\usepackage{epsfig}
\usepackage{graphicx}
\usepackage{amsmath}
\usepackage{amssymb}
\usepackage{bbm}
\usepackage{amsbsy}
\usepackage{slashed}

\usepackage{xcolor}

\definecolor{mbcol}{rgb}{1,0,1}

\usepackage[normalem]{ulem}  
\begin{document}

\title{Chiral Symmetry Restoration and Deconfinement  in the  Contact Interaction Model of Quarks with a  Parallel Electric and Magnetic Fields}
\author{Aftab Ahmad}
\affiliation{Institute of Physics and Electronics, Gomal University, 29220, D.I. Khan, K.P.K., Pakistan.}

\begin{abstract}
We study the impact of steady, homogeneous, and external parallel electric and magnetic field strength ($eE\parallel eB$), on the chiral symmetry breaking-restoration and confinement-deconfinement phase transitions. We also sketch the phase diagram of quantum chromodynamics (QCD) at finite temperature $T$ and in the presence of background fields. Our unified formalism for this study is based on the  Schwinger-Dyson equations, symmetry preserving vector-vector contact interaction model of quarks, and the proper time regularization scheme. At $T=0$, in the purely magnetic case ($eE\rightarrow 0$), we observe the well known magnetic catalysis effect. On the other hand, in the pure electric field background ($eB\rightarrow 0$), the electric field tends to restore the chiral symmetry and deconfinement above the pseudo-critical electric field $eE^{\chi, C}_c$. In the presence of both $eE$  and $eB$: we find the magnetic catalysis effect in the particular region where $eB$ dominates over $eE$, whereas, we observe the chiral inhibition (or electric chiral rotation) effect, when $eE$ stand over $eB$.  At finite $T$, in the pure electric field case,  the phenomenon of inverse electric catalysis appears to exist in our model.  On the other hand for pure magnetic field background,  we notice the magnetic catalysis effect in the mean-field approximation and inverse magnetic catalysis with $eB$-dependent coupling. The combined effect of both $eE$ and $eB$ on the pseudo-critical $T^{\chi, C}_c$ yields the inverse electromagnetic catalysis, with and without $eB-$dependent effective coupling of the model. Our findings are satisfactory in agreement with already predicted  results by lattice simulations and other reliable effective models of QCD.
\end{abstract}

\pacs{12.38.-t  
12.38.Aw ,11.10.Wx, 25.75.Nq, 98.62.En, 11.30.Rd}

\maketitle
\section{Introduction}
Dynamical chiral symmetry breaking and confinement are the two fundamental properties of the non-perturbative quantum chromodynamics (QCD). At zero or low temperature $T$, the fundamental degrees of freedoms of the hadronic matter are the low energy hadrons. At high temperature $T$ or density $\mu$, the dynamical chiral symmetry restored and deconfinement into another phase of hadronic matter with quarks and gluons becomes the new degrees of freedom, known as quark-gluon plasma (QGP).  Such a phase transition expected to happens in the early universe after a few micro-second of the big bang, and experimentally observed in heavy-ion collisions at
Large Hadron Collider (LHC) in CERN and Relativistic Heavy Ion Collider (RHIC) at Brook Heaven National Laboratory (BNL). Also, when the hadronic matter is subjected to an external electromagnetic field background, it yields a tremendous impact on phase transitions.  It is well known that at $T=0$, in pure magnetic case, the strong magnetic field tends to strengthen the formation of quark anti-quark condensate and the system remains in chiral symmetry broken phase even at the  high magnetic field strength $eB$, this phenomenon is known as the \textit{magnetic catalysis} (MC)~\cite{Klevansky:1988yw, Suganuma:1990nn, Klimenko:1990rh, Klevansky:1992qe, Krive:1992xh, Gusynin:1994re, Gusynin:1994va, Gusynin:1994xp}. It was examined in earlier studies that at finite $T$, the pseudo-critical temperature $T^{\chi, C}_c $, of the chiral symmetry restoration and deconfinement increases with the increase of $eB$ and hence, the magnetic catalysis percieved too  at finite $T$ \cite{Klevansky:1988yw, Suganuma:1990nn, Klimenko:1992ch, Klevansky:1992qe}. In the recent few years, the lattice QCD simulation~\cite{Bali:2012av, Bali:2013esa} predicted that at finite $T$, the magnetic field suppresses the formation of quark-antiquark condensate and tends to restore the chiral symmetry near the pseudo-critical temperature $T^{\chi, C}_c $. As a result, the $T^{\chi, C}_c $ decrease with the increase of $eB$, such a phenomenon is said to be~\textit{ the inverse magnetic catalysis} (IMC). 
This phenomenon is confirmed and supported by effective models of low energy QCD~\cite{Costa:2015bza, Ferrer:2014qka, Ayala:2015bgv, Andersen:2014oaa, Mueller:2015fka, Ahmad:2016iez, Farias:2016gmy} as well as in holographic QCD models\cite{He:2020fdi}.\\ %It is understood by realizing that being closer together, quark and antiquark pairs are reaching the asymptotic freedom regime faster by reducing the interaction strength as the intensity of the magnetic field increases~\cite{ayalaetal}. 
In pure electric case, and at $T=0$, the situation is quite different from that of pure magnetic field background.  The strong
electric field  opposes the formation of a  quark-antiquark condensate and 
thus tends to
restore the chiral symmetry, i.e., the electric field anti-screens the strong interaction, such a phenomenon is known as the  chiral electric inhibition effect~\cite{Klevansky:1988yw, Suganuma:1990nn, Klimenko:1992ch,Klevansky:1992qe, Cao:2015dya,Tavares:2018poq,Ruggieri:2016lrn}, or the chiral electric rotation effect~\cite{Wang:2017pje}. The nature of chiral phase transitions is of second order in the chiral limit, while cross-over when the bare quark mass is taken into account. At finite temperature, it is well understood that the pseudo-critical temperature $T^{\chi, C}_c $  decreases with an increase of electric field strength $e\vec{E}$, this is known as the  \textit{inverse electric catatalysis} (IEC)~\cite{Ruggieri:2016lrn, Tavares:2019mvq}. 
The study of the influence of the electric field on the chiral phase transitions is equally important as well as the effect of the magnetic field,  in both theoretical and experimental perspectives. 
Experimentally, in heavy-ion collisions, the electric and magnetic fields generated having the same order of magnitude ($\sim 10^{18}$ to $10^{20}$Gauss)~\cite{Bzdak:2011yy, Deng:2012pc, Bloczynski:2012en, Bloczynski:2013mca} in the event-by-event collisions using Au $+$ Au at  RHIC-BNL, and in a non-central heavy-ion collision of  Pb $+$ Pb in ALICE-LHC. Moreover, some interesting anomalous effects, such as  the chiral magnetic effect
%\footnote{It was
%shown that a magnetic field in the presence of imbalanced chirality induces a current along the magnetic field.
%Again, as a result, a positive charge is separated from a negative charge along the magnetic field. This is called the
%“Chiral Magnetic Effect”}
\cite{Kharzeev:2007jp,Fukushima:2008xe}, 
the chiral electric separation effect
%\footnote{ In the presence of both vector and axial charge densities, the electric field can induce an axial current along its direction and thus cause chirality separation} 
~\cite{Huang:2013iia, Jiang:2014ura}, the particle polarization effect~\cite{Karpenko:2016jyx, Xia:2018tes, Wei:2018zfb}, e.t.c., which may arise due to the generation of vector and/or axial currents in the presence strong electromagnetic fields, and needed to explore theoretically as well.
In recent few years, a  special case of considering the electric field strength parallel to the magnetic field strength ($eE\parallel eB$), paid much more attention to explore the above-mentioned phenomenon in the effective models of QCD~\cite{Ruggieri:2016lrn, Wang:2017pje, Wang:2018gmj, Cao:2020pjq}. One of the major reason is that the parallel electric and magnetic field plays an important and prominent role in many heavy-ion collision experiments~\citep{Li:2014bha, Zhao:2019ybo}.\\         
Keeping in view,  the above facts and findings, our motivation and aim are to understand the dynamical chiral symmetry breaking-restoration and confinement-deconfinement transitions, in the presence of strong and uniform external electromagnetic field. Our unified framework based on the Schwinger-Dyson equations (SDE) in the  rainbow-ladder truncation, the symmetry preserving  confining vector-vector contact interaction model 
(CI)\cite{Roberts:2011wy},
and the Schwinger proper time regularization scheme.  In the present work, we use the quark-antiquark condensate $\left\langle \psi \bar{\psi} \right\rangle$, an order parameters for the chiral symmetry breaking-restoration and  the confinement length scale~\cite{Ahmad:2016iez,Wang:2013wk} for the confinement-deconfinement transitions.   It is to be noted that the chiral symmetry restoration and deconfinement occur simultaneously in this model~\cite{Marquez:2015bca, Ahmad:2016iez}. \\ % This work is an extension to our previous work, wherein the pure magnetic case and at finite $T$,  we observed the phenomenon of magnetic catalysis (in the mean-field approximation) and inverse magnetic catalysis (with magnetic field dependent coupling) through contact interaction model~\cite{Ahmad:2016iez}. Here we are interested to include both the electric and magnetic fields.   
This article is organized as follows. In Sec. II, we present the general formalism and the contact interaction model at zero temperature and in the absence of background fields. In Sec.III, We discuss the  gap equation at zero temperature, and in the presence of parallel electric and magnetic fields in Sec.IV.  We draw the phase diagram at finite temperature and in the presence of parallel electric and magnetic fields in Sec. V.  In the last Sec.VI, we discuss the summary and perspectives.

\section{General Formalism and Contact Interaction Model} \label{section-2}
We  begin with  the Schwinger-Dyson's equations (SDE)  for  dressed-quark propagator $S_f$,  is given by:
\begin{eqnarray}
S^{-1}_f(p)&=&i\gamma \cdot p + m_f \nonumber\\
&&\hspace{-1.5cm} 
+\int \frac{d^4k}{(2\pi)^4} g^2
 \Delta_{\mu\nu}(p-k)\frac{\lambda^a}{2}\gamma_\mu S_f(k)
\frac{\lambda^a}{2}\Gamma_\nu(p,k)\, .
\label{eqn:gap-QCD}
\end{eqnarray}
Here the subscript $f$ represents the two light quark flavors i.e., up (u) and down (d) quarks, $g$ is the coupling constant, and  $m_f$ is the current quark mass,  which may set equal to zero in the chiral limit.  The $\lambda^a$'s are the
usual Gell-Mann matrices, $\Gamma_\nu $  is the dressed quark-gluon vertex,  
$\Delta_{\mu\nu}$ is the gluon propagator.\\
In the literature, it is well known that  the  properties of low energy hadrons     
can be  reproduced by  assuming that the interaction among the quarks, not
 via mass-less vector boson exchange, but instead through a   symmetry preserving four-fermions vector-vector contact interaction (CI), 
%In a series of previous articles, it has been shown
%that at zero temperature, the static properties of low
%energy mesons and baryons can be faithfully reproduced
%by assuming that quarks interact not via massless vector-boson
%exchange, but instead through asymmetry preserving
with a finite
gluon mass~\cite{GutierrezGuerrero:2010md,Roberts:2011wy,Roberts:2011cf,Roberts:2010rn,Chen:2012qr}
\begin{eqnarray}
  g^2 \Delta_{\mu \nu}(k) &=& \delta_{\mu \nu} \frac{4 \pi
  \alpha_{\rm ir}}{m_G^2} \equiv \delta_{\mu \nu} \alpha_{\rm eff}\,\label{eqn:CImodel} 
 \end{eqnarray}
where $\alpha_{\rm
ir}=0.93\pi $ is the infrared enhanced interaction strength parameter, $m_G=800$ MeV is the gluon mass scale~\cite{Boucaud:2011ug}.  In the CI  model,  for a small value of $\alpha_{\rm ir}$ and a large value of gluon mass scale $m_G$, there must be a critical value of  
$\alpha_{\rm eff}$,  above this critical value the chiral symmetry is broken and below this,  there is less chance for the generation of the dynamical mass. The d-dimensional (arbitrary space-time dimensions) dependence of this effective coupling and its critical value for the chiral symmetry breaking,  using an iterative method, in the superstrong regime, has been studied in detail in Ref.~\cite{Ahmad:2018grh}.\\
The CI model Eq.~(\ref{eqn:CImodel}) along with the choice of ranibow -ladder truncation $\Gamma_\nu(p,k)=\gamma_\nu$, form the kernel of  the quark SDE,  Eq.~(\ref{eqn:gap-QCD}), which bring the dressed-quark propagator  into a very simple form~\cite{Ahmad:2015cgh}:
\begin{equation}
S_f^{-1}(p)=i\gamma\cdot p+M_f\,.\label{eqn:gapCI}
\end{equation}
It is because the wavefunction renormalization trivially
tends to unity in this case, and the quark mass function $M_f$
become momentum independent ~:
\begin{equation}
M_f=m_f+\frac{4\alpha_{\rm eff}}{3}\int^\Lambda \frac{d^4k}{(2\pi)^4} {\rm Tr}[S_f(k)]\;.\label{gapNJL}
\end{equation}
In this truncation, the  quark-anitquark condensate  is given by
\begin{equation}
-\left\langle \psi \bar{\psi} \right\rangle= N_c\int^\Lambda \frac{d^4k}{(2\pi)^4} {\rm Tr}[S_f(k)]\;,\label{gapNJLc}
\end{equation} 
with $N_c=3$, are the number of colors. The form of our gap equation Eq.~(\ref{gapNJL}), is very much  similar to the NJL model gap equation  accept the coupling parameter  $ \alpha_{\rm eff}=\frac{9}{2} G $
~\cite{Marquez:2015bca}.\\ 
Further simplification of Eq.(\ref{gapNJL}) yields the following gap equation.
\begin{equation}
M_f= m_{f}+ \frac{16\alpha_{\rm eff}}{3} \int^\Lambda \frac{d^4q}{(2\pi)^4}\frac{M_f}{k^2+M_f^2}\;,\label{eqn:gapNJLd}
\end{equation}
where $M_f$ is the dynamical mass and the symbol $\int^\Lambda$ stresses the need to regularize the integrals. Using $d^4 k= (1/2) k^2 dk^2 \sin ^2 \theta  d\theta \sin \phi d \phi d\psi $, performing the trivial regular integration's and using the variable $s=k^2$, the above expression reduces to:
\begin{equation}
M_f=m_{f}+\frac{ \alpha_{\rm eff} M_f }{8\pi^2}\int^{\infty }_{0}
ds\frac{s}{s+M_f^2} \, \label{eqn:PTRab}
\end{equation}
Obivously, the  integral in Eq.~(\ref{eqn:PTRab}) is not convergent,  so we need to regularize it through  proper-time regularization scheme~\cite{Schwinger:1951nm,Klevansky:1988yw, Suganuma:1990nn,Klevansky:1992qe} . In this  scheme, we take the exponentiation of the
denominator of the integrand, by introducing  an additional  infrared cutoff, beside the usual ultraviolet cut-off, normally used in NJL model studies. In this way, the confinement is  implemented through an infrared cut-off
~\cite{Ebert:1996vx}. Upon adopting this  scheme,   
the quadratic and logarithmic divergences removes and the axial-vector Ward-Takahashi identity~\cite{Ward:1950xp,Takahashi:1957xn} is satisfied.
From  Eq.~(\ref{eqn:PTRab}), 
the denominator of the integrand is given by

\begin{eqnarray}
\frac{1}{s+M^{2}_{f}}&=&\int^{\infty }_{0} d\tau {\rm e}^{-\tau(s+M^{2}_{f})}
\rightarrow 
\int^{\tau_{\text{ir}}^2}_{\tau_{\text{uv}}^2} d\tau {\rm
e}^{-\tau(s+M^{2}_{f})} \nonumber \\
&=&\frac{ {\rm e}^{-\tau_{\text{uv}}^2(s+M^{2}_{f})}-{\rm e}^{-\tau_{\text{ir}}^2(s+M^{2}_{f})}}{s+M^{2}_{f}}
. \label{eqn:PTRe}
\end{eqnarray}

Here, $\tau_{\text{uv}}^{-1}=\Lambda_{\text{uv}}$ is ultra-violet regulator, which play the dynamical role and set the scale for all dimensional quantities.
The $\tau_{\text{ir}}^{-1}=\Lambda_{\text{ir}}$ stand for the  infra-red regulator, whose non zero value implements confinement by ensuring the absence of quarks production thresholds ~\cite{Roberts:2007jh}.  Hence,  $\tau_{\text{ir}}$ is corresponds to the confinement scale~\cite{Ahmad:2016iez}. From Eq.~(\ref{eqn:PTRe}), it is now clear that the location of the original pole is at $s=-M^2$, which is canceled by the numerator. In this way, we have removed the singularities, and thus the propagator is free from real as well as the complex pole, which is consistent with the definition of confinement i.e., ``an excitation
described by a pole-less propagator would never reach its
mass-shell''~\cite{Ebert:1996vx}. \\
After performing integration over `s', the gap equation is given by:

\begin{eqnarray}  
 M_f&=& m_{f} + \frac{M_f^3 \alpha_{\rm eff}}{8\pi^{2}}
  \Gamma(-1,\tau_{\text{uv}} M_{f}^2,\tau_{\text{ir}} M_{f}^2)\,,\label{eqn:const_mass_reg} 
\end{eqnarray}
\noindent where
\begin{equation}
\label{eqn:incomplete_gamma}
\Gamma (a, x_1,x_2)=\Gamma (a,x_1)-\Gamma(a,x_2)\,,
\end{equation}
with $\Gamma(a,x) = \int_x^{\infty} t^{\alpha-1} {\rm
e}^{-t} dt$, is the incomplete Gamma function. By using the parameters of 
 Ref.~\cite{GutierrezGuerrero:2010md}, i.e., $\alpha_{\rm eff}(0) = 5.739\cdot 10^{-5}~\mathrm{MeV^{-2}}$, $\tau_{\text{ir}} = (240~\mathrm{MeV})^{-1}$ and 
       $\tau_{\text{uv}} = (905~\mathrm{MeV})^{-1}$,   
with the bare quark masses $m_f=7$~MeV, we get dynamical mass $M_f=367 MeV$, 
 and  condensate $\langle\bar{u}u\rangle^{1/3} = \langle\bar{d}d\rangle^{1/3} = -243$~MeV.\\ In the next section, we discuss the gap equation in the presence external electromagnetic field  at zero temperature.

\section{Gap equation at $T=0$ and in the background of  Parallel $eE$ and $eB$ } 
\begin{figure}[t!]
\begin{center}
\includegraphics[width=0.48\textwidth]{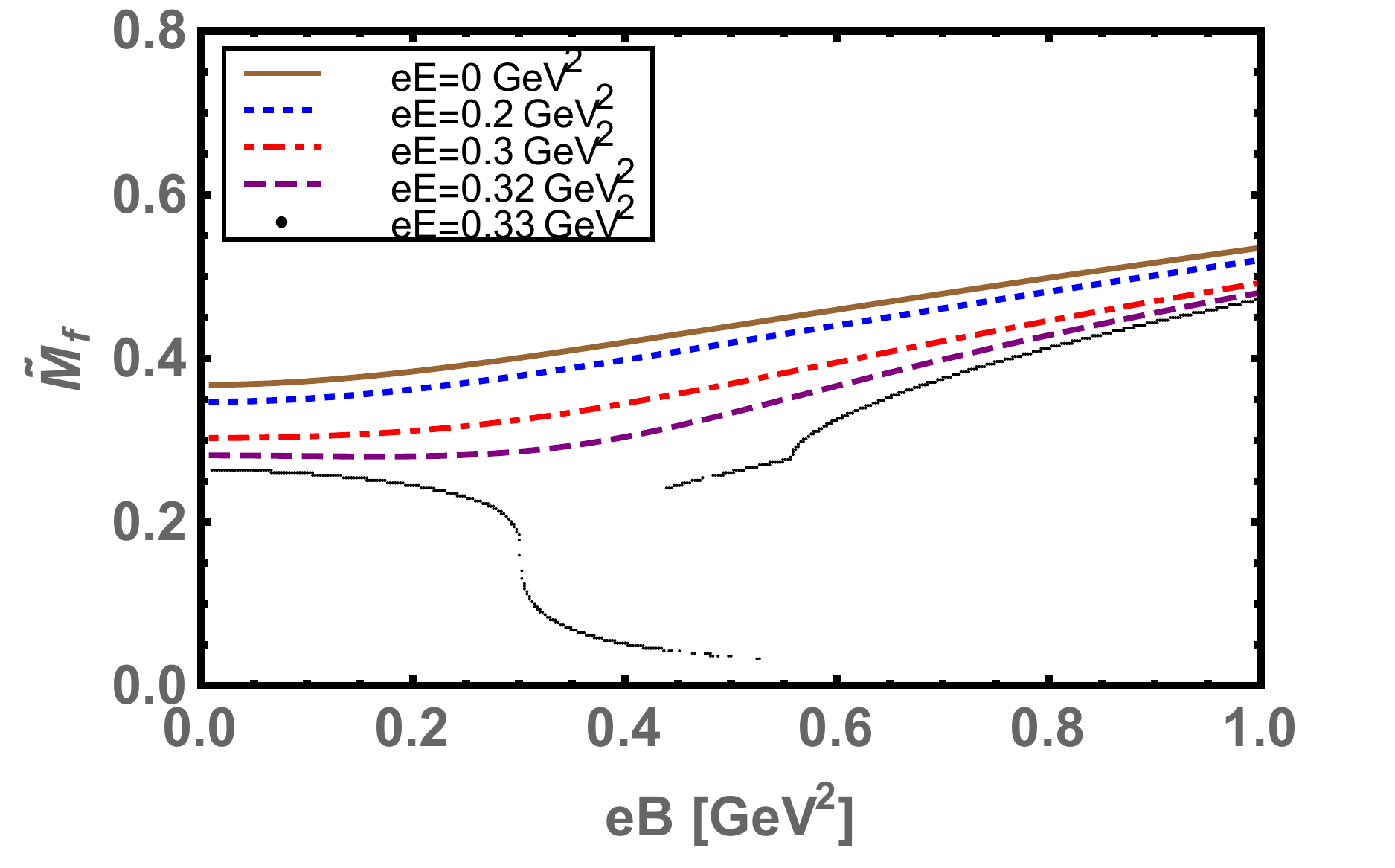}
\caption{ %Average Mass; \textit{Mid panel}: 
Behavior of the dynamically generated quark mass  as a function of  magnetic field  strength $eB$ with a bare quark mass 
$m_{f}=7$~MeV for several given values of $eE$.}
\label{Fig1}
\end{center}
\end{figure}

\begin{figure}[t!]
\begin{center}
\includegraphics[width=0.48\textwidth]{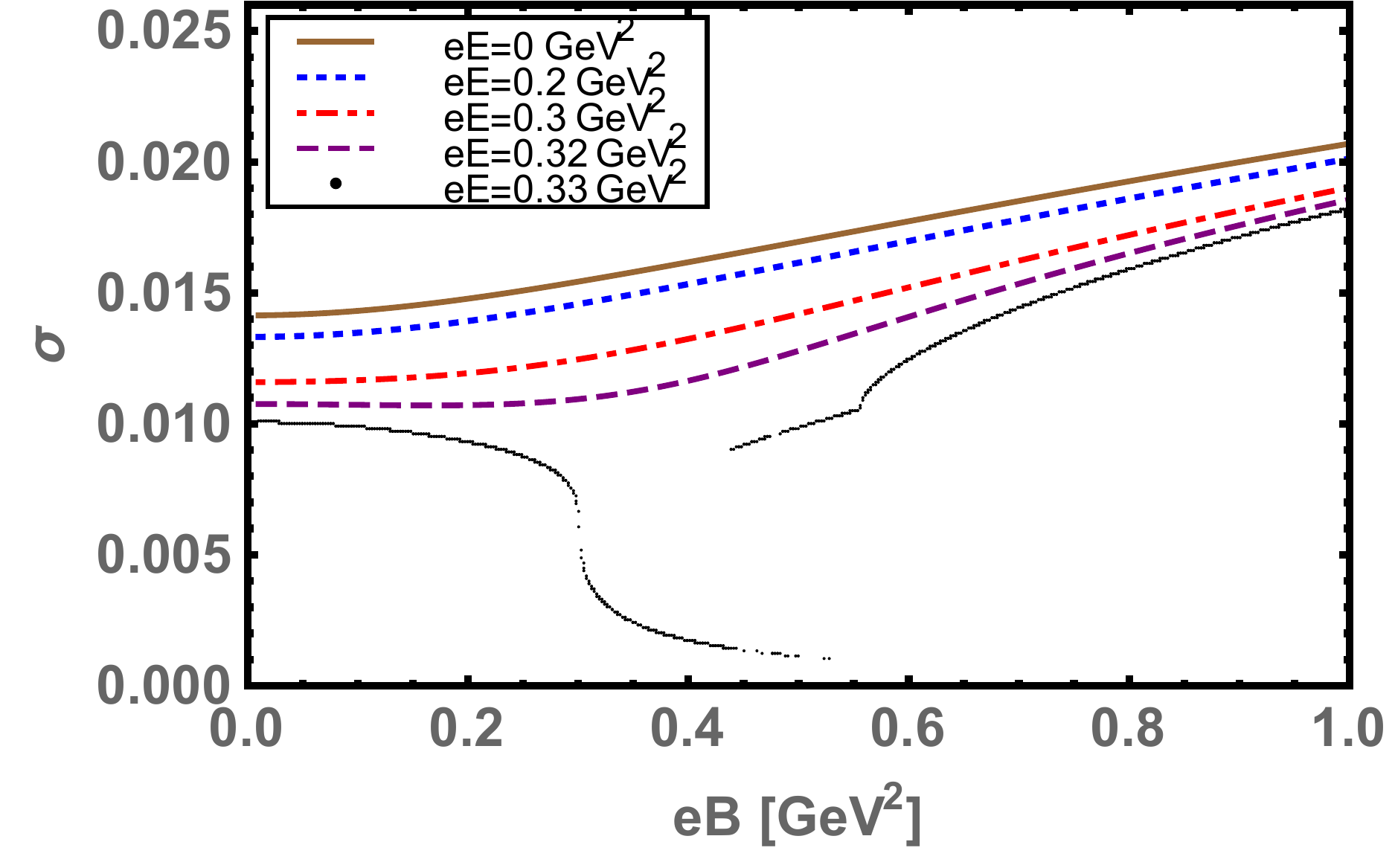}
\caption{ %Average Mass; \textit{Mid panel}: 
Quark-antiquark condensate  as a function of  magnetic field  strength $eB$ for several values of $eE$.}
\label{Fig2}
\end{center}
\end{figure}

\begin{figure}[t!]
\begin{center}
\includegraphics[width=0.48\textwidth]{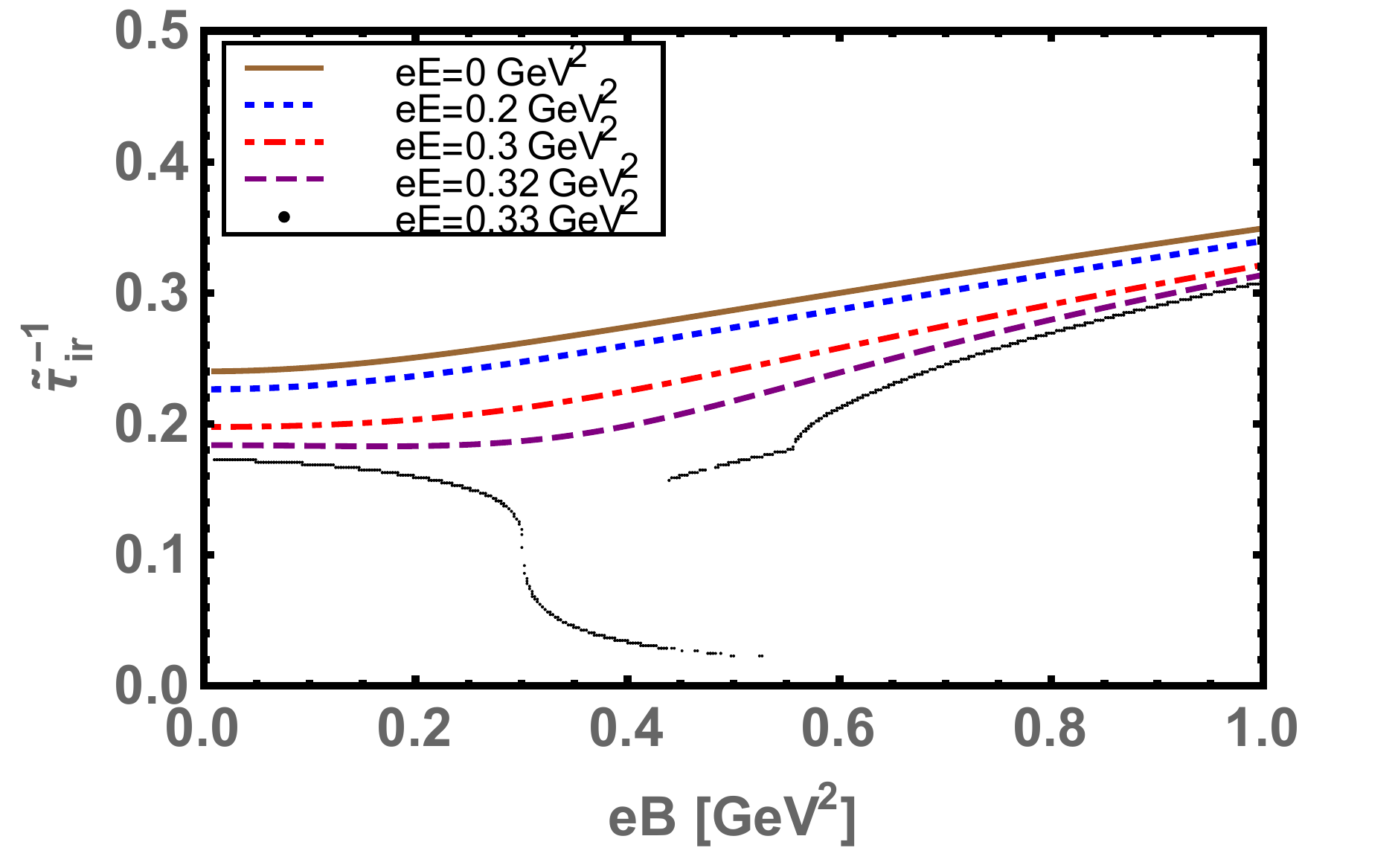}
\caption{ %Average Mass; \textit{Mid panel}: 
 The confinement length scale $\tilde{\tau}^{-1}_{ir}$ as a function of  magnetic field  strength $eB$ for several values of $eE$.}
\label{Fig3}
\end{center}
\end{figure}

\begin{figure}[t!]
\begin{center}
\includegraphics[width=0.48\textwidth]{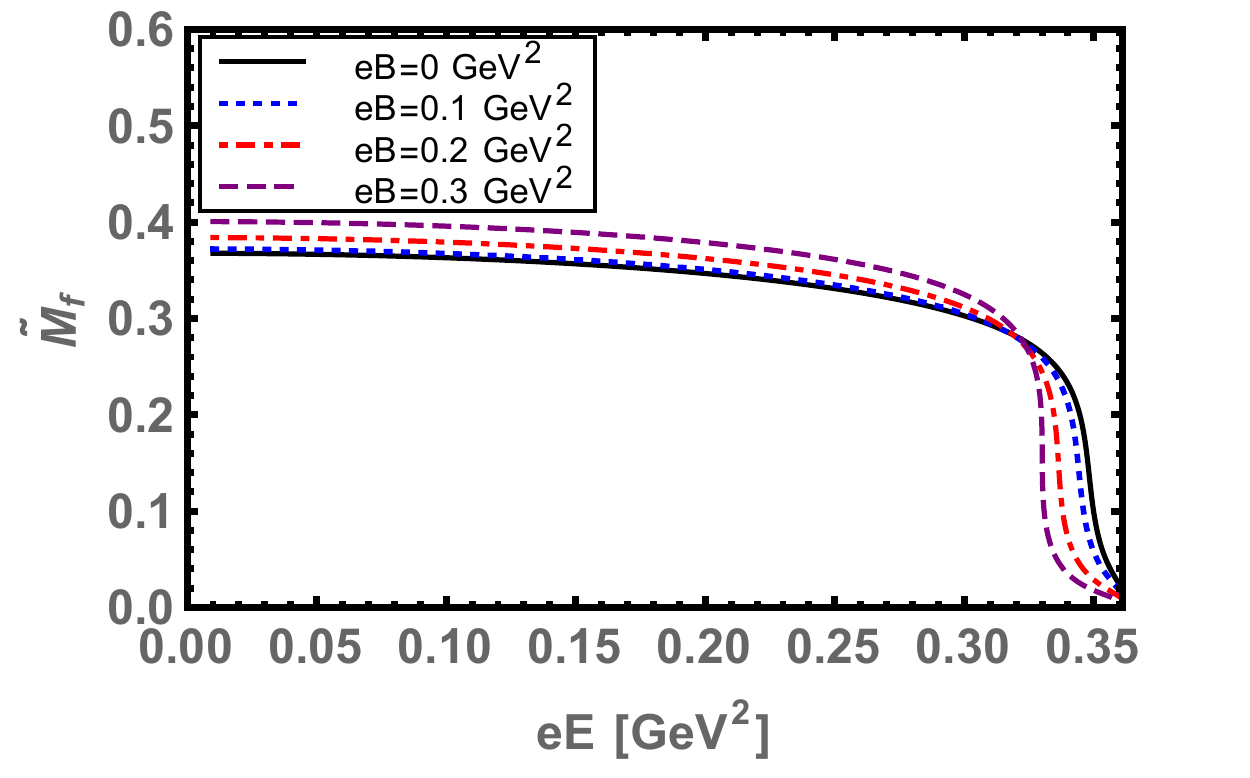}
\caption{ %Average Mass; \textit{Mid panel}: 
The  dynamical mass as a function of  electric field  strength $eE$  for several values of $eB$.}
\label{Fig4}
\end{center}
\end{figure}

\begin{figure}[t!]
\begin{center}
\includegraphics[width=0.48\textwidth]{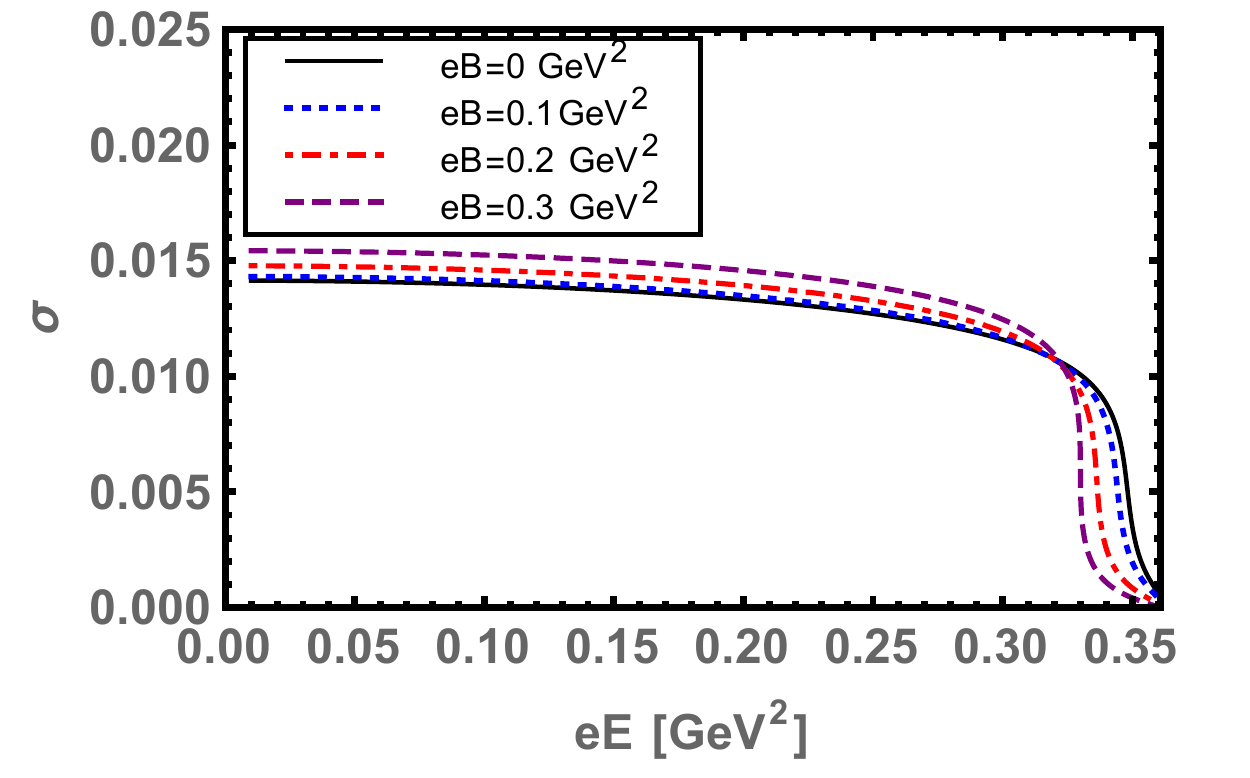}
\caption{ %Average Mass; \textit{Mid panel}: 
Quark-antiquark condensate $\sigma$ as a function of  electric field  strength $eE$ for several values of $eB$.}
\label{Fig5}
\end{center}
\end{figure}

\begin{figure}[t!]
\begin{center}
\includegraphics[width=0.48\textwidth]{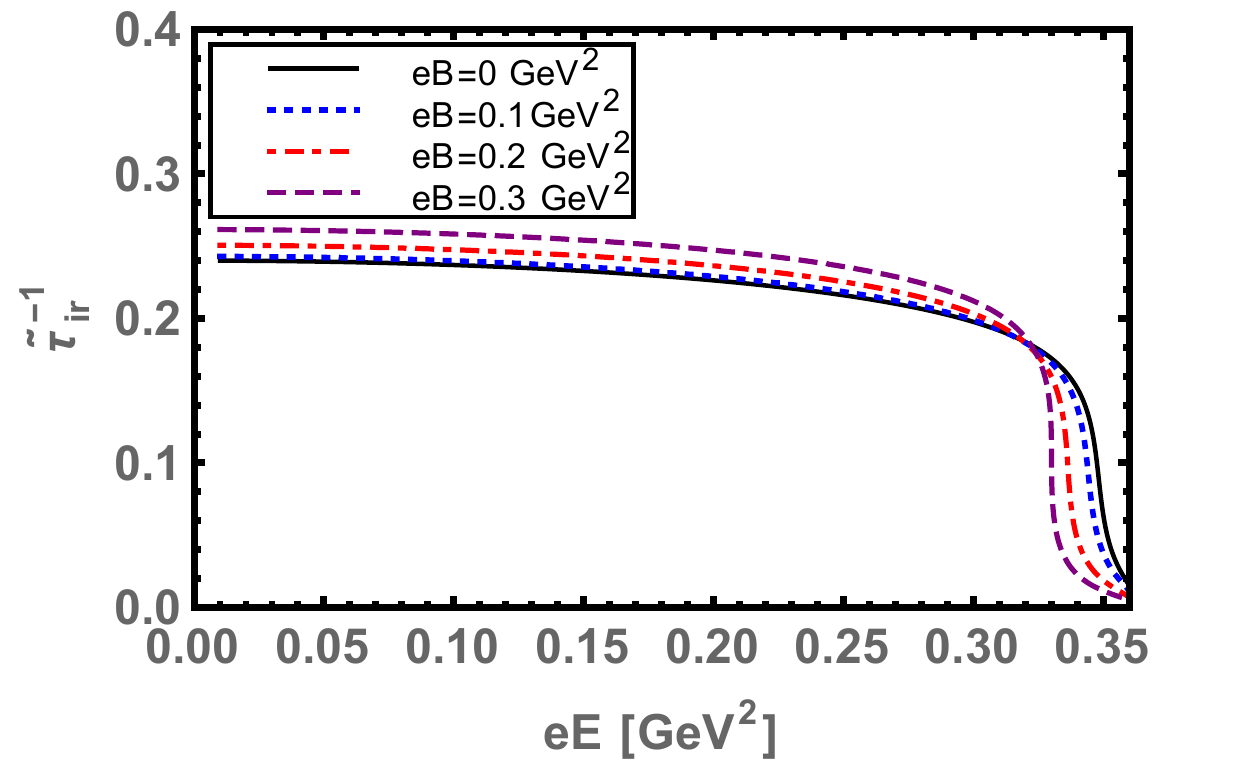}
\caption{ %Average Mass; \textit{Mid panel}: 
Confinement length scale $\tilde{\tau}^{-1}_{ir}$ as a function of  electric field  strength $eE$ for  a given several values of $eB$. }
\label{Fig6}
\end{center}
\end{figure}

\begin{figure}[t!]
\begin{center}
\includegraphics[width=0.48\textwidth]{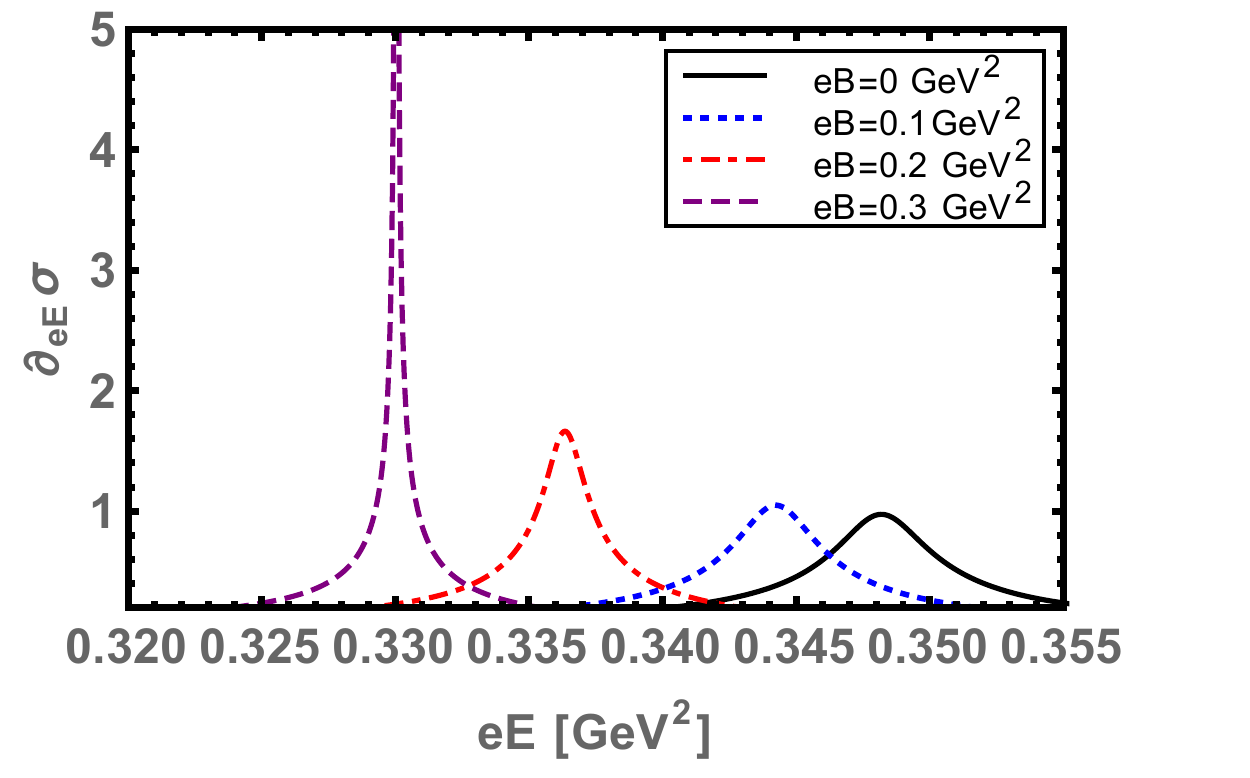}
\caption{ %Average Mass; \textit{Mid panel}: 
The electric gradient of the quark-antiquark condensate $\partial_{eE}\sigma $, as a function of $eE$ for fixed values of $eB$.  At particular fixed value of $eB^{\chi}_c=0.3$ GeV$^2$, the electric gradient  $\partial_{eE} \sigma$ diverges at $eE^{\chi}_{c,}\approx0.33$ GeV$^2$, and above the first order phase transition occurs.}
\label{Fig7}
\end{center}
\end{figure}

\begin{figure}[t!]
\begin{center}
\includegraphics[width=0.48\textwidth]{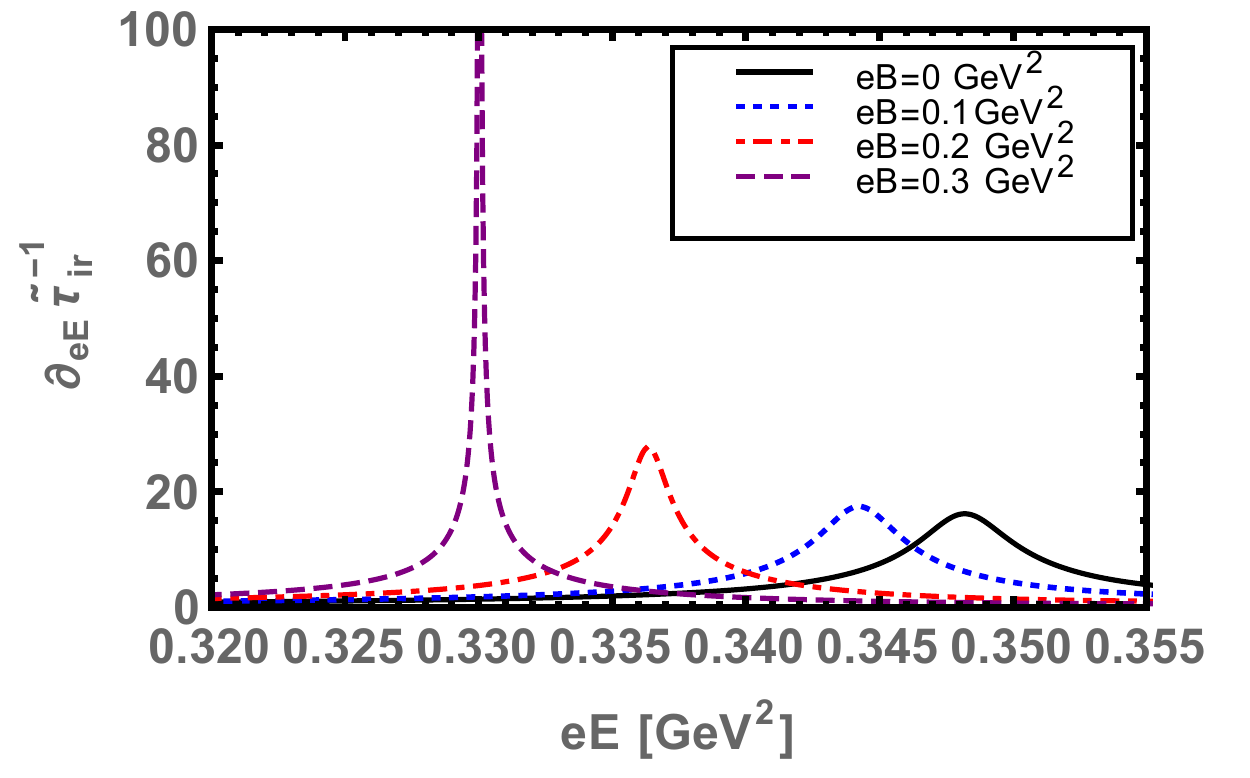}
\caption{ %Average Mass; \textit{Mid panel}: 
The electric gradient of the confinement length scale $\partial _{eE} \tilde{\tau}^{-1}_{ir}$ as a function of $eE$, for different fixed value of $eB$. At particular fixed value of $eB^{C}_c=0.3$GeV$^2$, the electric gradient of the confinement scale diverges at $eE^{C}_{c,}\approx0.33$ GeV$^2$.}
\label{Fig8}
\end{center}
\end{figure}

\begin{figure}[t!]
\begin{center}
\includegraphics[width=0.48\textwidth]{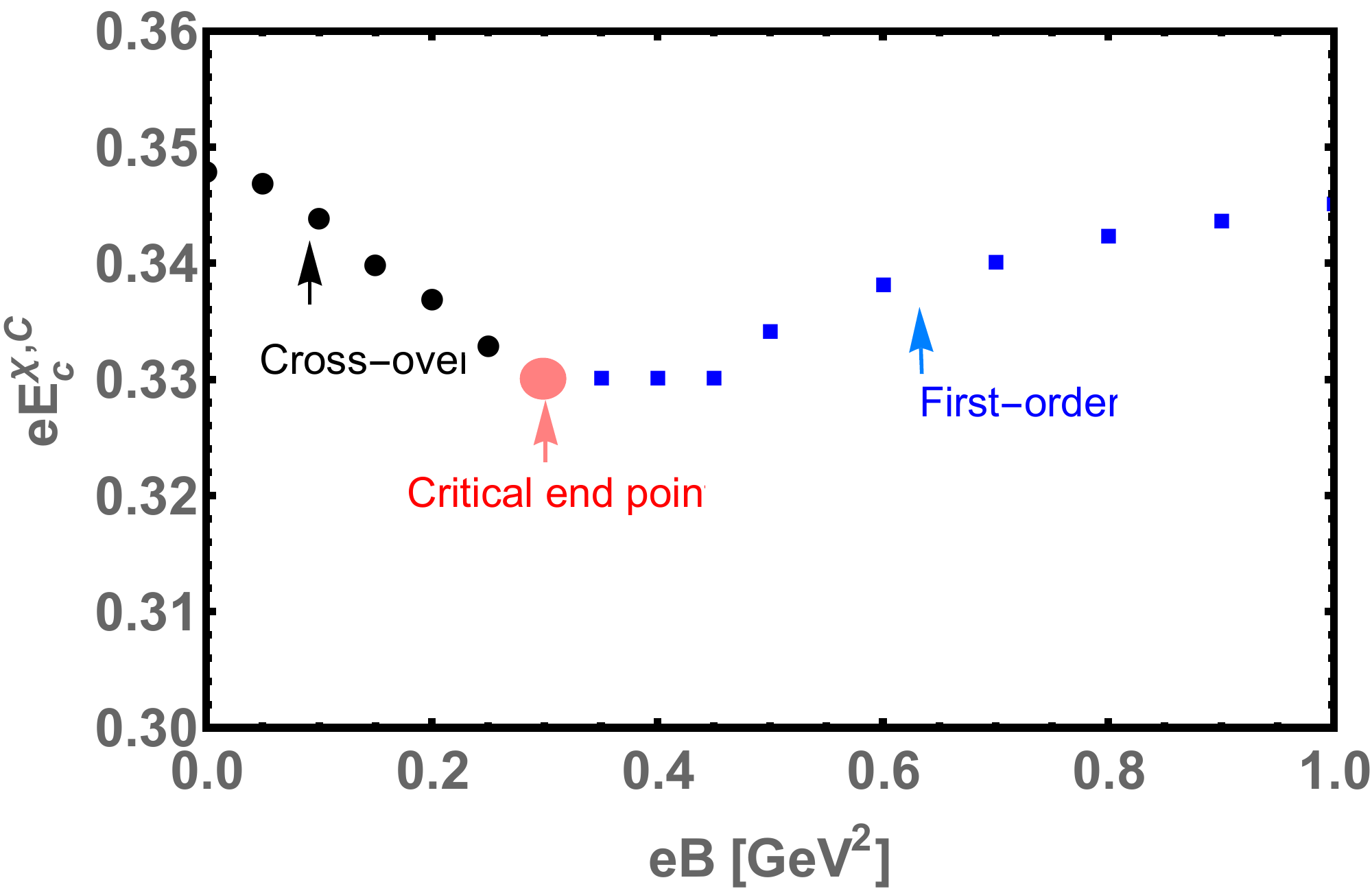}
\caption{ %Average Mass; \textit{Mid panel}: 
The phase diagram of chiral symmetry and confinement for $eE^{C,\chi}_{c}$ vs $eB$:  The $eE=eE^{C,\chi}_{c}$, is obtained from the inflection  points  of the  electric gradient of the condensate $\partial_{eE}\sigma$ and the confining length scale
$\partial _{eE} \tilde{\tau}^{-1}_{ir}$.}
\label{Fig9}
\end{center}
\end{figure}

In this section, we study the gap equation in the presence of a uniform and homogeneous electromagnetic field with  $eE\parallel eB$, at zero temperature. In QCD Lagrangian, the interaction  with  parallel electromagnetic field $A^{ext}_\mu$  embedded in the  covariant derivative,
\begin{eqnarray}
 D_\mu=\partial_\mu -iQ_f A_\mu^{\rm ext}, \label{em1} 
\end{eqnarray}
with $Q_{f}=(q_u=+2/3 ,q_d=-1/3)e$ is refers to the electric charges of $u$ and $d$-quark respectively. We use  the symmetric gauge  vector potential $A^{ext}_\mu=(i E z, 0,-B x, 0)$ in Euclidean space, with both electric and magnetic field are chosen  along the  z-axis. The gap equation in the presence of parallel electromagnetic field continues to be of the form Eq.~(\ref{gapNJL}), where $S_f(k)$ is now  dressed 
with parallel background fields, i.e., $ {S_f}(k) \rightarrow\tilde{S_f}(k)$.
The $\tilde{S_f}(k)$, in Schwinger proper time  representation~\cite{Schwinger:1951nm,Klevansky:1988yw, Suganuma:1990nn,Klevansky:1992qe}, in the presence of  parallel magnetic field in Euclidean space~\cite{Cao:2015dya,Wang:2017pje}, can be written as:  
\begin{eqnarray}
\tilde{S_f}(k)&=&\int^{\infty}_{0} d\tau {\rm e}^{-\tau \bigg( M_{f}^{2}+(k_{4}^{2}+k_{3}^{2}) \frac{{\rm tan}(|Q_{f}E| \tau)}{|Q_{f}E|\tau}+(k_{1}^{2}+k_{2}^{2})\frac{{\rm tanh}(|Q_{f}B|\tau)}{|q_{f}B|\tau}\bigg)}
\nonumber\\&& \times \bigg[-\gamma^4 k_4+M_f+{\rm tan}(|Q_{f}E| \tau )
\bigg(\gamma^4 k_3-\gamma^3 k_4\bigg)\nonumber\\&&  -i {\rm tanh}(|Q_{f}B\tau |)
\bigg(\gamma^1 k_2-\gamma^2 k_1\bigg) \bigg]\nonumber\\&& \times
\bigg[1-i{\rm tanh}(|Q_{f}B|\tau) {\rm tan}(|Q_{f}E|\tau )\gamma^5
\nonumber\\&&  -i{\rm tanh}(|Q_{f}B |\tau)\gamma^1 \gamma^2+{\rm tan}(|Q_{f}E|\tau)\gamma^4 \gamma^3 \bigg], \label{em2}
\end{eqnarray} 
where the  electric and magnetic field coupled to the  momentum co-ordinates $k_{4}$, $k_{3}$ and  $k_{1}$, $k_{2}$, respectively.  
Now, taking the \textit{trace} of Eq.~(\ref{em2}), and introducing both infrared and ultraviolet cut-offs, the gap equation Eq.(~\ref{gapNJL}), in the presence of constant parallel electromagnetic field is given by 
\begin{eqnarray}
\tilde{M_f}&=& m_{f}+ \frac{ 16\alpha_{\rm eff}}{3}\sum_{f=u,d} \int^{\tilde{\tau}^{2}_{ir}}_{\tau^{2}_{uv}} d\tau \tilde{M}_f {\rm e}^{-\tau \tilde{M}_{f}^{2}} \nonumber\\
&&\times\int\frac{d^2 k_{a}}{(2\pi)^2}\frac{d^{2} k_{b}}{(2\pi)^{2}} {\rm e}^{-\tau ( k_{a}^{2} \frac{{\rm tan}(|Q_{f}E| \tau)}{|Q_{f}E|\tau}+k^{2}_{b}\frac{{\rm tanh}(|Q_{f}B|\tau)}{|Q_{f}B|\tau}},  \label{em3}
\end{eqnarray}
with $k^{2}_{a}=k_{4}^{2}+k_{3}^{2}$ and $k^{2}_{b}=k_{1}^{2}+k_{2}^{2}$. The Lorentz structure preserves in this case. 
The confining scale now is 
\begin{eqnarray}
\tilde{\tau}_{ir}=\tau_{ir}\frac{M_f}{\tilde{M}_f},\label{em4}
\end{eqnarray}
where $M_f$ is the the dynamical mass and $\tilde{M}_f$ is the electromagnetic field dependent dynamical mass. In the chiral limit, $\tilde{\tau}_{ir}\to\infty$ (or equivalently $\tilde{\tau}_{ir}\to 0$ ), at  the chiral symmetry restoration region. In the presence of finite current quark masses,  it allows poles to be develop in the propagator, which ensures the coincidence between confinement and chiral symmetry transitions~\cite{Ahmad:2016iez}. 
After integration over $k$  the  gap equation Eq.~(\ref{em3}), can be written as  % at a finite temperature under the influence of magnetic field  

\begin{eqnarray}
\tilde{M}_f &=& m_{f}+ \frac{\alpha_{\rm eff}}{3\pi^2}\sum_{f=u,d} \int^{\bar{\tau}^{2}_{ir}}_{\tau^{2}_{uv}} d\tau \tilde{M}_f {\rm e}^{-\tau \tilde{M}_{f}^{2}}
\nonumber\\&& \times\bigg[\frac{|Q_{f}E|}{{\rm tan}(|Q_{f}E|\tau)} \frac{|Q_{f}B|}{{\rm tanh}(|Q_{f}B|\tau)}\bigg].  \label{em5}
\end{eqnarray}
The gap equation for pure electric  field can be obtained by setting $eB\to 0$, while for  pure magnetic field,  $eE\to 0$. 
The quark-antiquark condensate in the presence of background fields is of the form:

\begin{eqnarray}
-\left\langle \psi \bar{\psi} \right\rangle &=& \frac{3}{4\pi^2}\sum_{f=u,d} \int^{\bar{\tau}^{2}_{ir}}_{\tau^{2}_{uv}} d\tau \tilde{M}_f {\rm e}^{-\tau \tilde{M}_{f}^{2}}
\nonumber\\&& \times \bigg[\frac{|Q_{f}E|}{{\rm tan}(|Q_{f}E|\tau)} \frac{|Q_{f}B|}{{\rm tanh}(|Q_{f}B|\tau)}\bigg]  \label{em6},
\end{eqnarray}
In this present  work,  we use  two flavors $f=2$, i.e., $u$- and $d$-quarks. We take the  current quark masses $m_u=m_d=7$ MeV, and hence the iso-spin   symmetry is preserved in this way.   As we know that, the response of electromagnetic field is different for $u$ and $d$-quarks, it is because of heaving  different electric charges. In the present scenario, we ignore the charge
difference of $u$ and $d$-quarks and solve Eq.~(\ref{em5}) and Eq.~(\ref{em6}) for an
average common charge $\overline{Q_f}=(|Q_u|+|Q_d|)/2$~\cite{Klevansky:1988yw,Klevansky:1992qe}.\\ 
The numerical solution of the gap equation Eq.~(\ref{em5}) with finite current quark mass $m_f=7$ MeV, as a function of  $eB$ for  fixed values of  $eE$, is plotted in the  Fig.~\ref{Fig1}.
In pure magnetic case ($eE\rightarrow0$), the dynamical  mass $\tilde{M_f}$  monotonically increase with increase of  $eB$,  and thus,  ensures the phenomenon of \textit{magnetic catalysis}.  The increase in  $\tilde{M_f}$ as a function of  $eB$ reduces in magnitude upon varying $eE$, from its smaller to larger given values. We find that at $eE\geq0.33$ GeV $^2$, The dynamical mass $\tilde{M_f}$, show the de Haas-van Alphen oscillatory  type behavior\citep{deHaas1930}: it remain constant for small $eB$, then  monotonically decrease with the increase of $eB$, and suddenly jump down to its lower values  in the region $eB=[0.3-0.54]$~GeV$^2$, where the  chiral symmetry partially restored via first order phase transition, and above this region it increases again. It may be due to the fact that 
there is a strong competition occurs between parallel $eB$ and $eE$, i.e., in one hand $eB$  enhances the mass function, while on the other hand $eE$ suppresses it. The same behavior is explores and argued in the effective model of QCD, see for example in Ref.~\cite{Wang:2017pje,Cao:2015xja}.\\ The behavior of the quark-antiquark condensate $\sigma=-\left\langle \psi \bar{\psi} \right\rangle $, i.e., Eq.~(\ref{em6}) as a function of  $eB$, for various  given values of $eE$, is shown in the Fig.~\ref{Fig2}. 
For $eE\rightarrow0$, the magnetic field strength facilitates the formation of quark-antiquark condensate, and hence the system remains in the enhanced dynamical chiral symmetry broken phase . The condensate $\sigma$ behave in a similar fashion as the dynamical mass  $\tilde{M_f}$:  at a given values of  $eE\geq0.33$ GeV$^2$,  the evolution is suppressed in the region $eB=[0.3-0.54]$~GeV$^2$, the chiral symmetry partially restored, the nature of transition become first-order and above that region, it is again broken and enhanced with $eB$. The pseudo-critical values of the fields, at which the chiral symmetry partially restored and first-order phase transition occurred are at $eB_c=0.3$ GeV$^2$ and  $eE_c=0.33$ GeV$^2$. Such a values of electric or magnetic fields strength is large enough in terms of what is usually generated during the heavy ion collisions, but  may  relevant to the astronomical objects: like neuron stars, magnetars, e.t.c. The confinement parameter $\tilde{\tau}^{-1}_{ir}$ as a function of $eB$ for different fixed values of $eE$, depicted in the Fig.~\ref{Fig3}. It shows, the similar behavior as $\tilde{M_f}$ and $\sigma$.  We find the same pseudo-critical fields $eB_c=0.3$ GeV$^2$ and  $eE_c=0.33$ GeV$^2$ for the confinement transition, as we find in the  case of chiral symmetry breaking.\\  
In the following, we now discuss the variation of $\tilde{M_f}$, $\sigma$ and $\tilde{\tau}^{-1}_{ir}$ as a function of $eE$, in pure electric field as well as for several non-zero values of $eB$.    
In Fig.~\ref{Fig4}, Fig.~\ref{Fig5} and Fig.~\ref{Fig6},  we shows the behaviors of all the three parameters as a function of $eE$, for different fixed values of $eB\geq0$. 
In the  pure electric field limit ($eB\rightarrow0$), all the three parameters decrease monotonically with the increase of $eE$ and antisreening effect appears at a pseudo-critical field strength $eE^{\chi,C}_c$, where the chiral symmetry partially restore and deconfinement transitions occurs. The nature of the transition here is a smooth cross-over.  We thus, observe the the chiral rotation effect or chiral elecric inhibiion effect in the contact interaction model, as already predicted by other effective models of QCD \cite{Klevansky:1988yw,Suganuma:1990nn,Cao:2015dya, Ruggieri:2016lrn, Tavares:2018poq, Wang:2017pje}. For  non-zero $eB$, we find an interesting behaviors of all the three parameters as a function of $eE$. All three parameters, enhances for a given  smaller to larger values of  $eB$, except the region of chiral symmetry restoration and deconfinement, where all the parameters suppressed  by the higher value of $eB$. The pseudo-critical field strength $E^{\chi,C}_{c}$ decreases with the increase of $eB$, and at some pseudo-critical $eB_c\approx 0.3$~GeV$^2$, the transition changes from cross-over to first order. The nontrivial behaviors of all three parameters represents, the competition between magnetic catalysis
effect and  electric inhibition effect, both induced by $eB$ in the
presence of parallel $eE$~\cite{Wang:2017pje}. 
The magnitude of $ eE^{\chi,C}_{c}$,  at which the chiral symmetry partially restore and deconfinement takes place, is triggered from the inflection point of the electric gradient $\partial_{eE}\sigma$ and  $\partial _{eE} \tilde{\tau}^{-1}_{ir}$, as shown in the Fig.~\ref{Fig7} and Fig.~\ref{Fig8}, respectively. 
In pure electric case, the chiral symmetry restored and deconfinement transition occurs at $eE^{\chi,C}_{c}\approx0.348$ GeV$^2$. For several given values of $eB\neq0$, the  $eE^{\chi,C}_{c}$ decreases with the increase of $eB$, we find a smooth  cross-over phase transition till at the pseudo-critical magnetic field strength $eB_c\approx0.3$ GeV$^2$ and above this value, the transition become first-order.  Now $eE^{\chi,C}_{c}$  remains constant in the region $eB\approx[0.3-0.54]$ GeV$^2$, and then increases with the larger values of $eB$ as shown in the Fig.~\ref{Fig9}, the same behavior is already demonstrated in~\citep{Wang:2017pje}.  We call the boundary point, where the cross-over phase transitions end and the first order phase transition start, as a critical end point, whose co-ordinates are at $(eB_{p}=0.3,eE_{p}=0.33)$ GeV $^2$. 
In the next section, we study the chiral symmetry breaking-restoration and confinement-deconfinement phase transition at finite temperature and in the presence of parallel electromagnetic  field background and at the end we sketch the phase diagrams.
\section{QCD phase diagram  in  the  presence of Parallel $eE$ and  $eB$} 
\begin{figure}[t!]
\begin{center}
\includegraphics[width=0.48\textwidth]{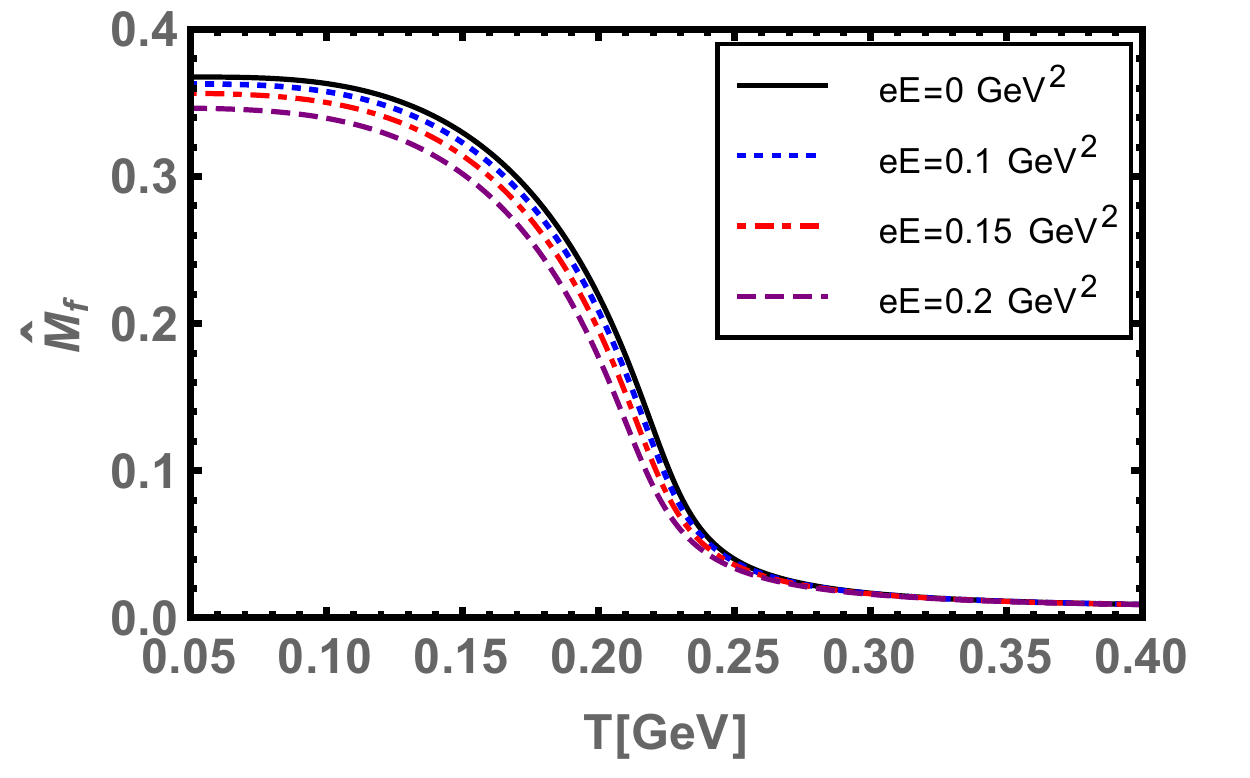}
\caption{ %Average Mass; \textit{Mid panel}: 
The dynamical mass as a function of  temperature  for various given values of electric field strength $eE$.}
\label{Fig10}
\end{center}
\end{figure}

\begin{figure}[t!]
\begin{center}
\includegraphics[width=0.48\textwidth]{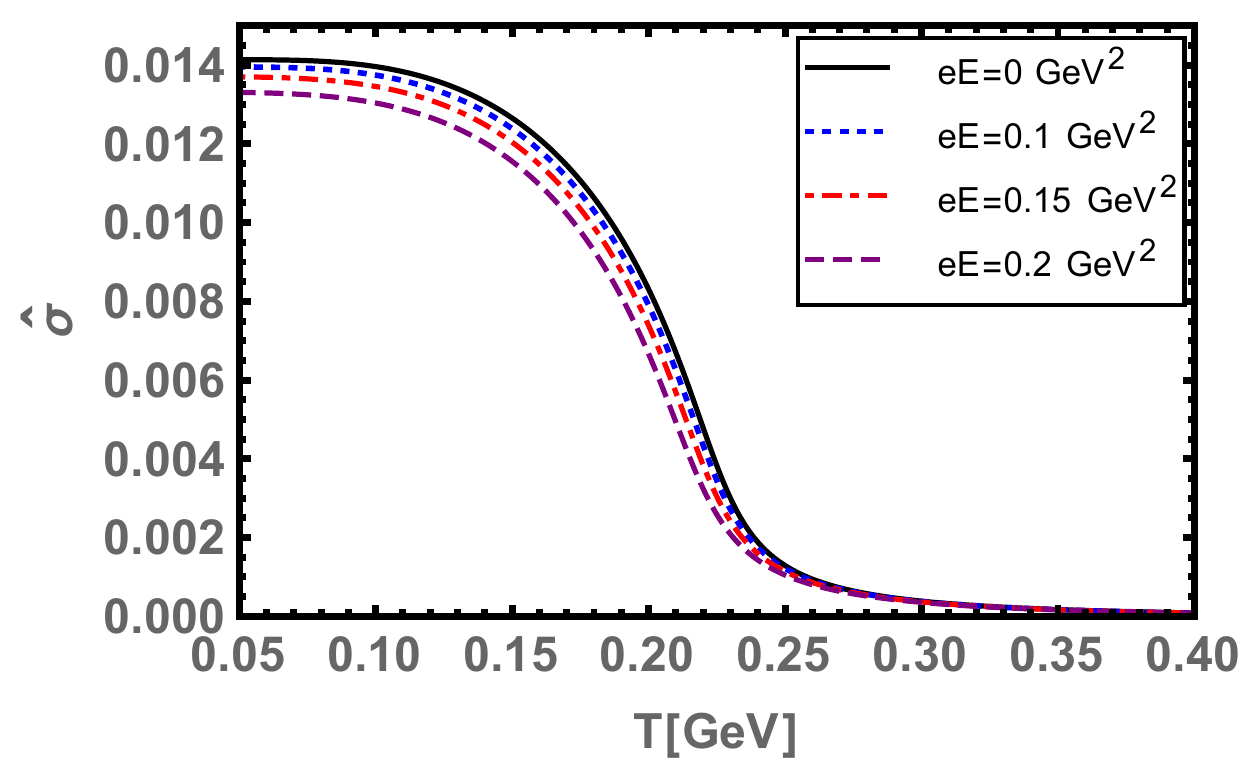}
\caption{ %Average Mass; \textit{Mid panel}: 
The quark-antiquark condensate $\widehat{\sigma}=-\widehat{\left\langle \psi \bar{\psi} \right\rangle}$ as a function of  temperature  for various given values of electric field strength $eE$.}
\label{Fig11}
\end{center}
\end{figure}

\begin{figure}[t!]
\begin{center}
\includegraphics[width=0.48\textwidth]{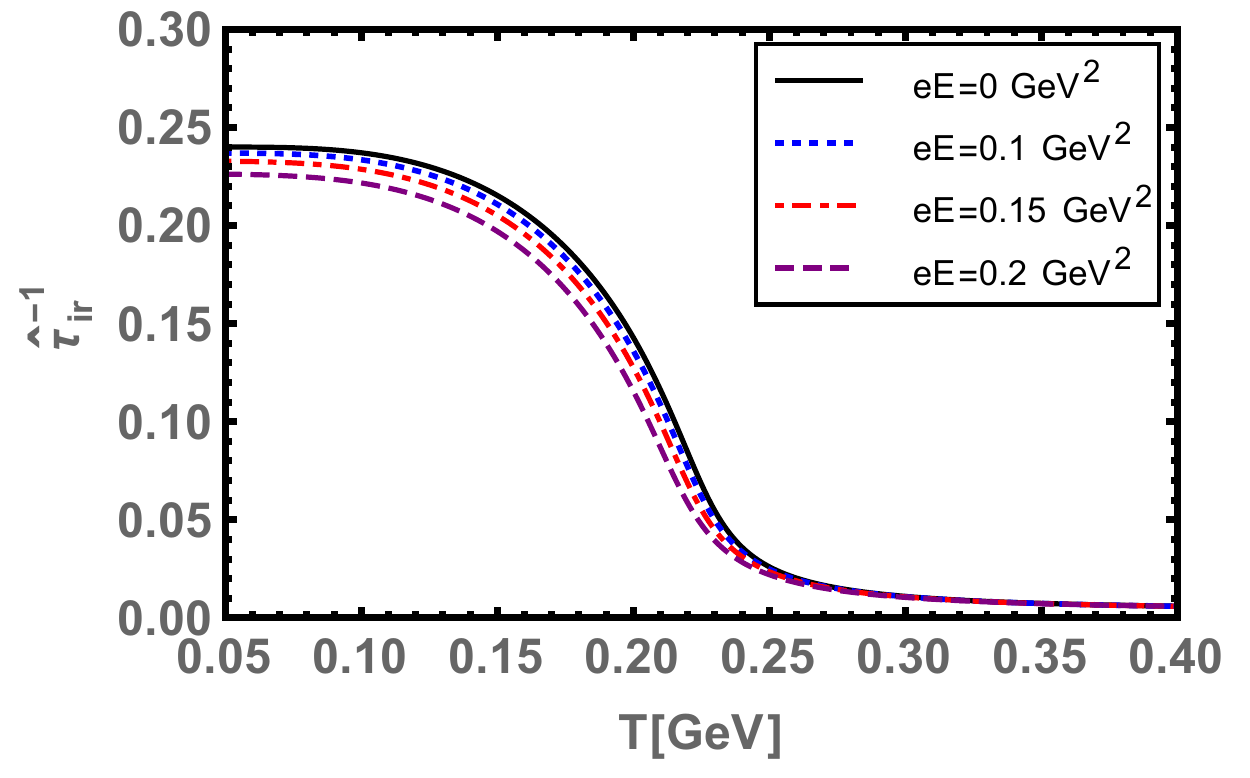}
\caption{ %Average Mass; \textit{Mid panel}: 
The behavior of confinement length scale  $\widehat{\tau}^{-1}_{ir}$ plotted as a function of temperature, for  different given values of  electric field  strength $eE$.}
\label{Fig12}
\end{center}
\end{figure}

\begin{figure}[t!]
\begin{center}
\includegraphics[width=0.48\textwidth]{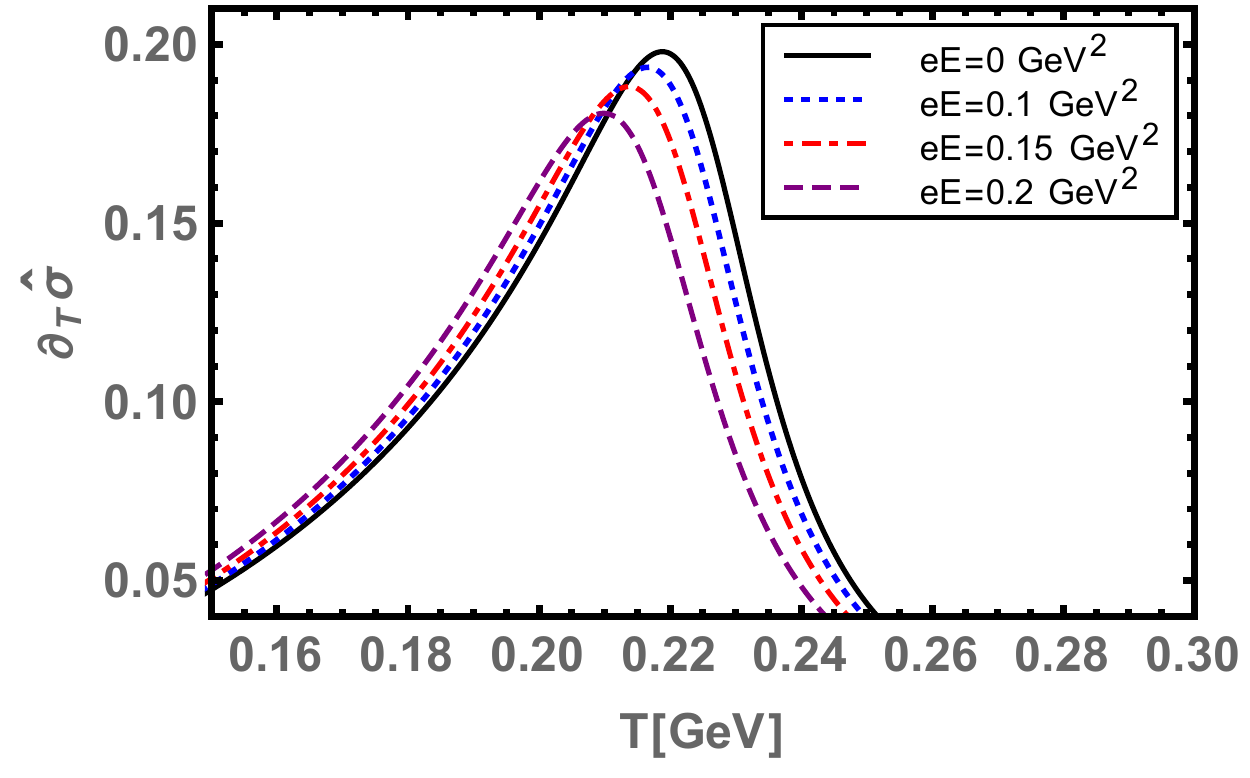}
\caption{ %Average Mass; \textit{Mid panel}: 
The thermal gradient of the quark-antiquark condensate $\partial _{T}\widehat{\sigma}$, plotted  as a function of temperature for several values of electric field $eE$. The peaks in the derivatives shifted towards low temperature region, for  small to large values of electric field strength. }
\label{Fig13}
\end{center}
\end{figure}

\begin{figure}[t!]
\begin{center}
\includegraphics[width=0.48\textwidth]{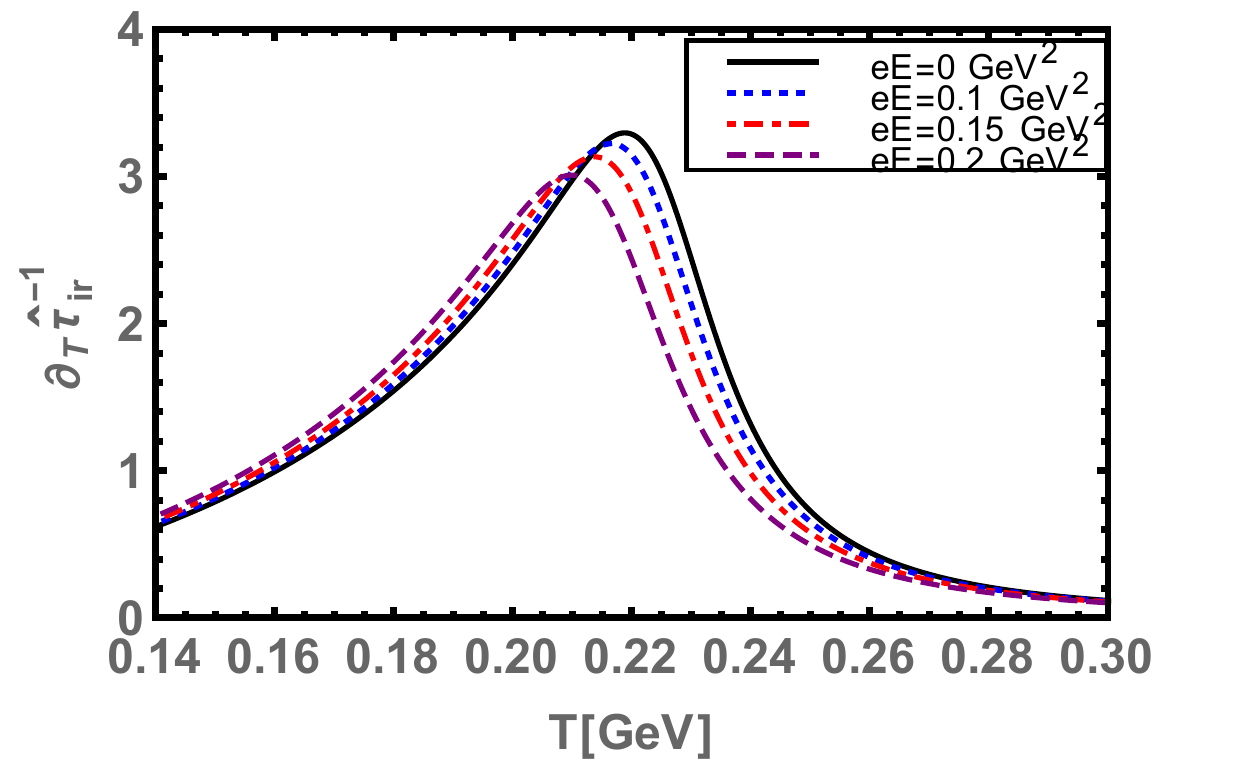}
\caption{ %Average Mass; \textit{Mid panel}: 
The thermal gradient of the confinement length scale $\partial _{T}\widehat{\tau}^{-1}_{ir}$, plotted  as a function of temperature for several values of electric field $eE$.}
\label{Fig14}
\end{center}
\end{figure}

\begin{figure}[t!]
\begin{center}
\includegraphics[width=0.48\textwidth]{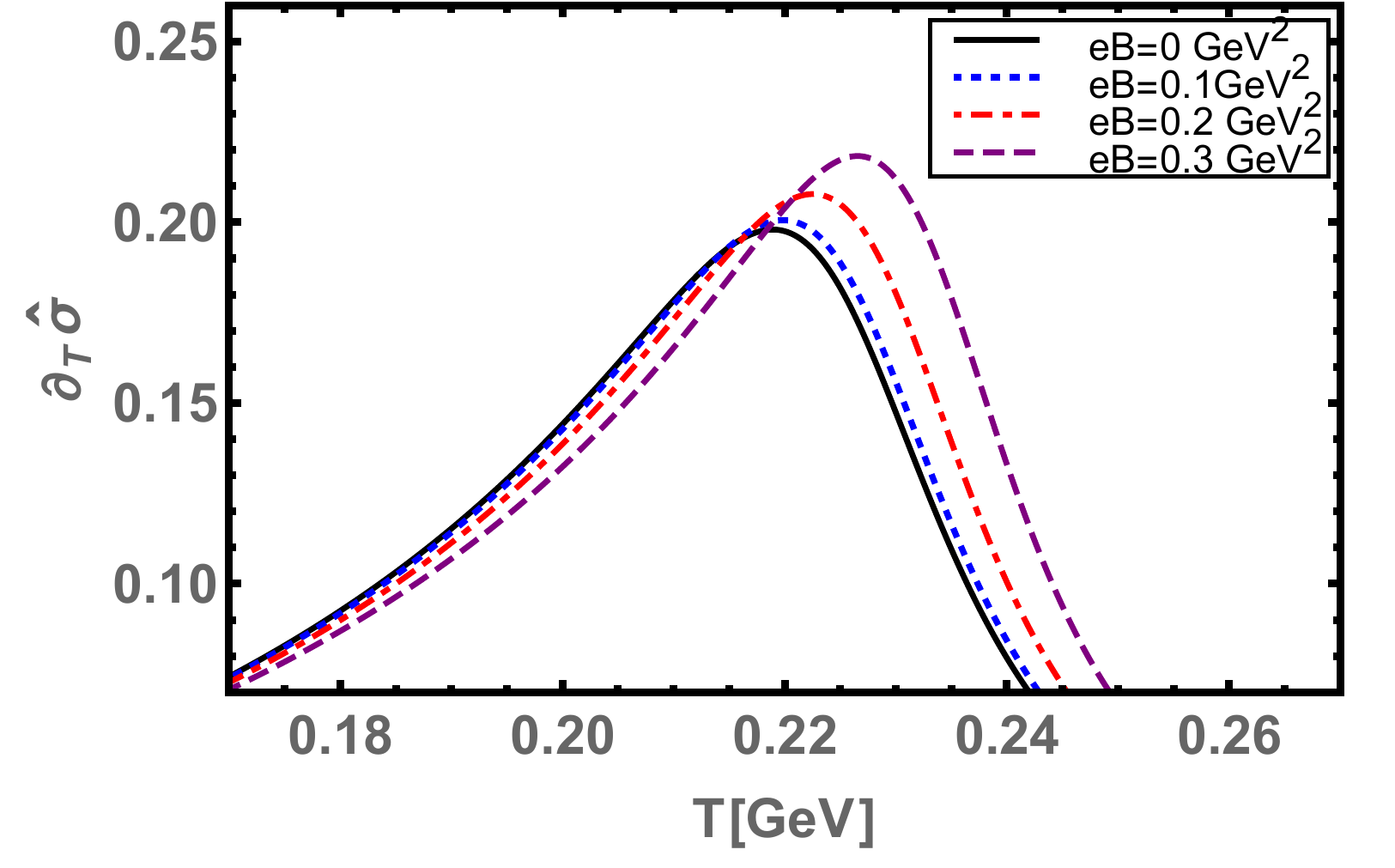}
\caption{ %Average Mass; \textit{Mid panel}: 
The thermal gradient of the quark-antiquark condensate $\partial _{T}\widehat{\sigma}$, plotted  as a function of temperature for several values of magnetic field  strength $eB$.  As it is clear from the plots that peaks shifting toward higher critical temperatures.}
\label{Fig15}
\end{center}
\end{figure}

\begin{figure}[t!]
\begin{center}
\includegraphics[width=0.48\textwidth]{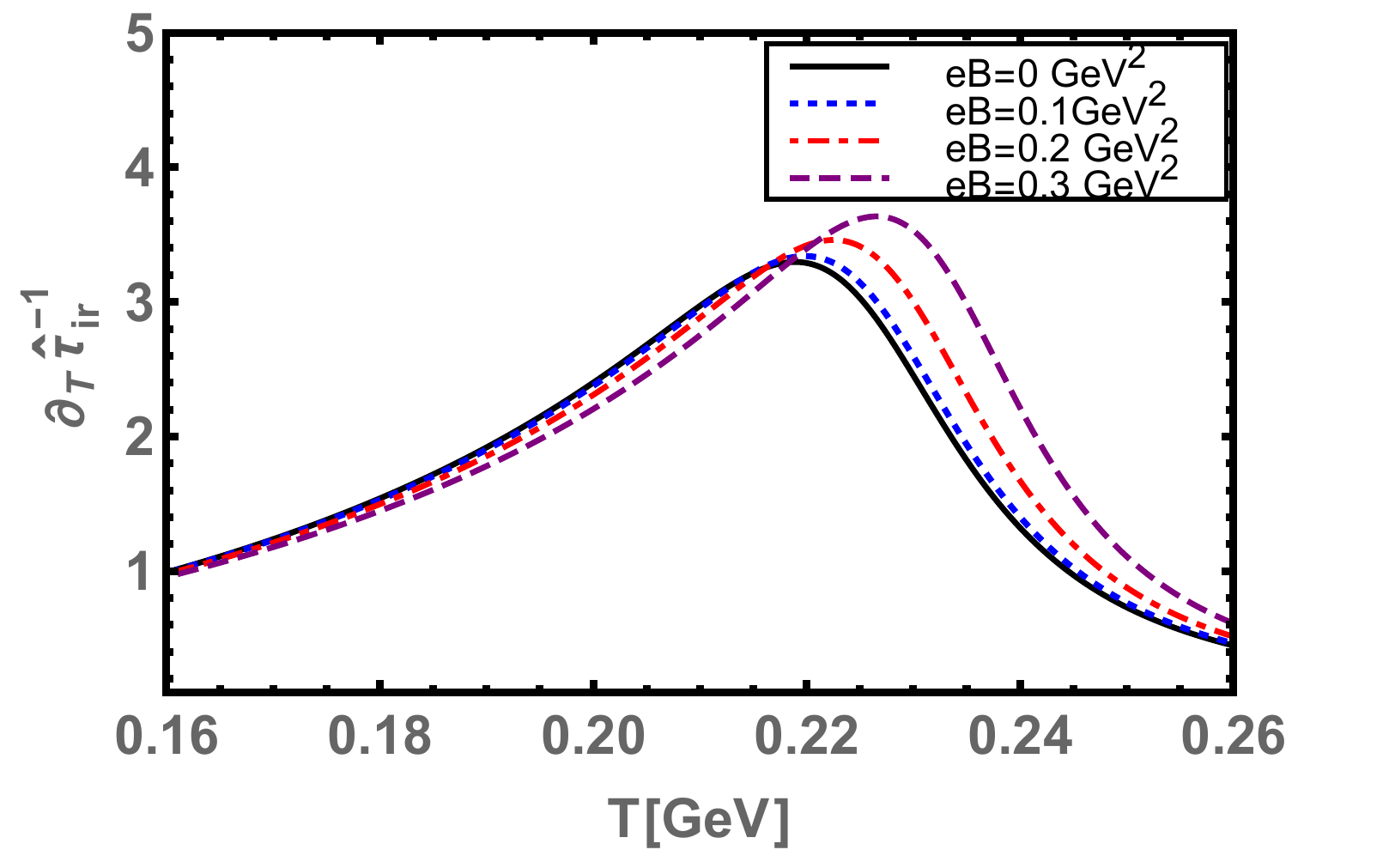}
\caption{ %Average Mass; \textit{Mid panel}: 
The thermal gradient of the confinement length scale $\partial _{T}\widehat{\tau}^{-1}_{ir}$, plotted  as a function of temperature for several values of magnetic field strength field $eB$.}
\label{Fig16}
\end{center}
\end{figure}

\begin{figure}[t!]
\begin{center}
\includegraphics[width=0.48\textwidth]{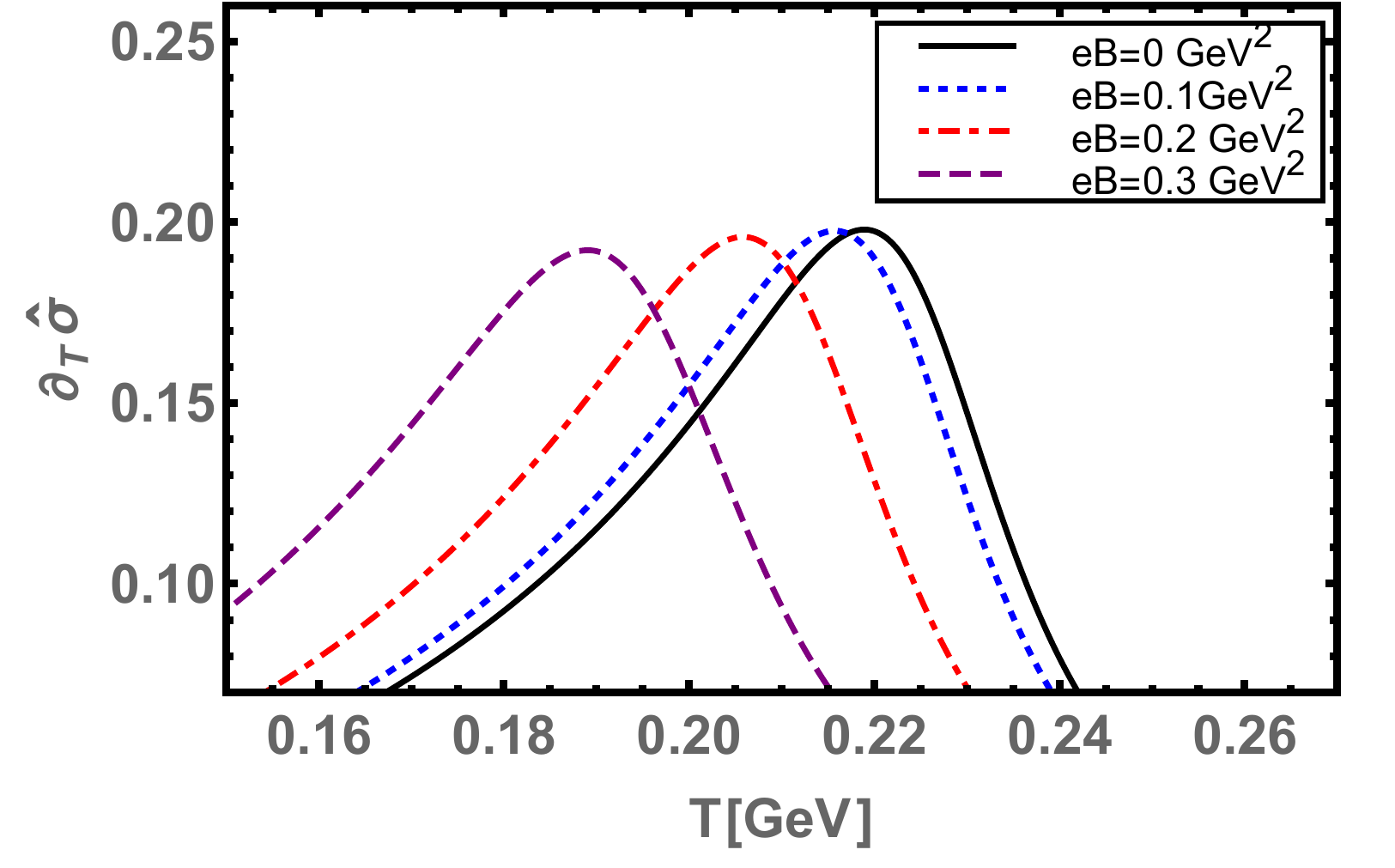}
\caption{ %Average Mass; \textit{Mid panel}: 
The thermal gradients of the condensate with magnetic field dependent coupling~Eq.~(\ref{emt6}), plotted  as a function of temperature, for several values of magnetic field strength. It shows the IMC effect.}
\label{Fig17}
\includegraphics[width=0.48\textwidth]{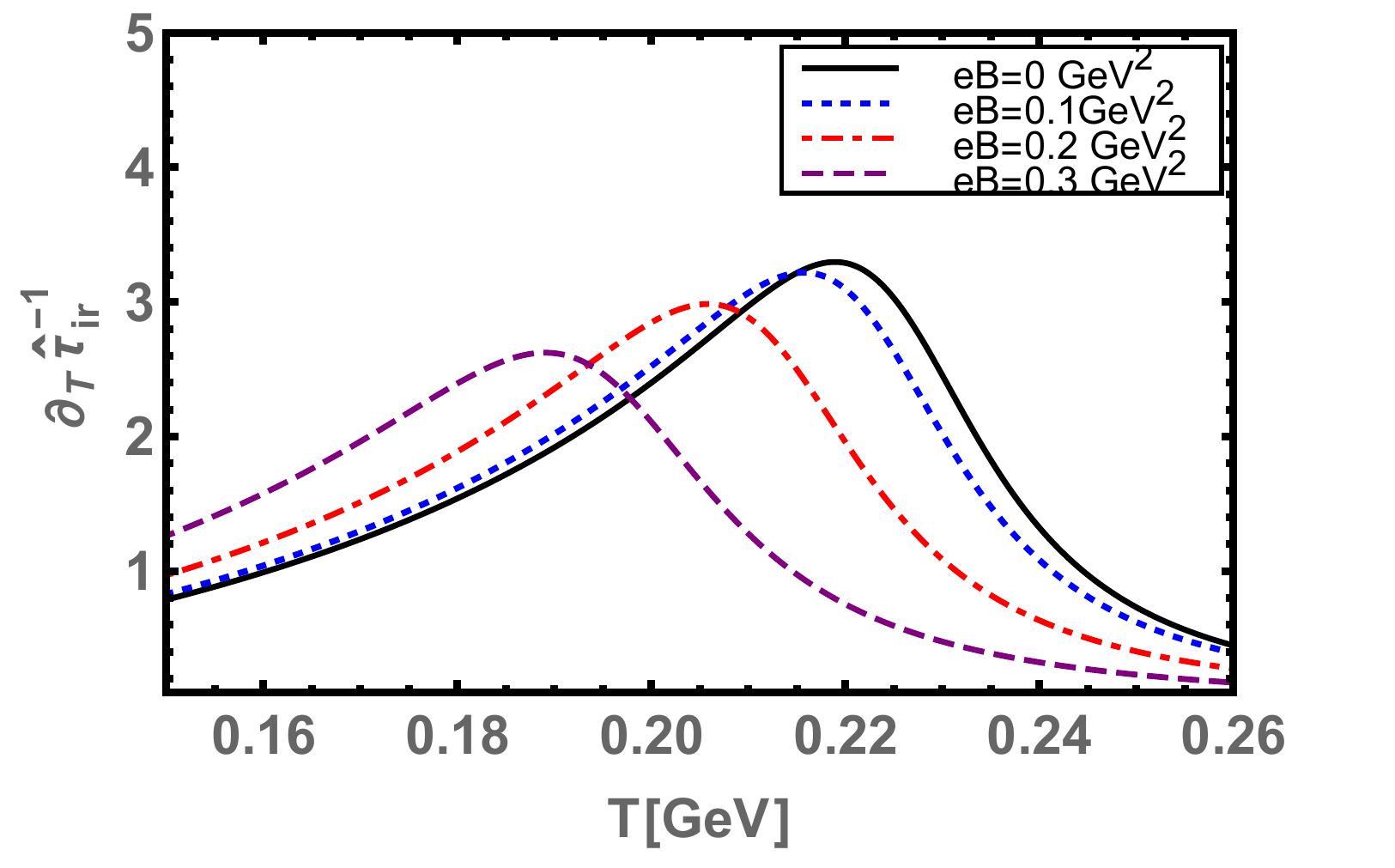}
\end{center}
\end{figure}

In this section, we explain how the dynamical mass, the condensate, and the confinement length scale, behave at finite temperature and in the presence of parallel electric and magnetic fields. We also explore the phenomenon of inverse electric catalysis (IEC), magnetic catalysis (MC), inverse magnetic catalysis (IMC), and the competition among them.\\ 
The finite temperature version of the gap equation Eq.~(\ref{em3}) in the presence of parallel electric and magnetic field can be obtained by adopting the standard convention for momentum integration
\begin{eqnarray}
\int\frac{d^4k}{(2\pi)^4} \rightarrow T \sum_{n} \int\frac{d^3k}{(2\pi)^3}, \label{emt1}
\end{eqnarray}
and  the four momentum $k\rightarrow(\omega_{n},\vec{k})$, with $\omega_n = (2n+1)\pi T$ are  the fermionic Matsubara frequencies. The Lorentz structure does not preserve anymore at finite temperature. By making the following replacements in Eq.~(\ref{em3}),
\begin{eqnarray}
\int\frac{d^2 k_{a}}{(2\pi)^2}\frac{d^{2} k_{b}}{(2\pi)^{2}}\rightarrow T \sum_{n}\int\frac{d^2 k_{3}}{(2\pi)}\frac{d^{2} k_{b}}{(2\pi)^{2}},\nonumber
\end{eqnarray}
$k^{2}_{a}\rightarrow \omega^{2}_{n}+k_{3}^{2}$ and  
$k^{2}_{b}\rightarrow k_{1}^{2}+k_{2}^{2}$,  we have 
\begin{eqnarray}
\widehat{M_f}&=& m_{f}+ \frac{ 16T\alpha_{\rm eff}}{3}\sum^{\infty}_{n=-\infty}\sum_{f=u,d}
 \int^{\widehat{\tau}^{2}_{ir}}_{\tau^{2}_{uv}} d\tau \widehat{M_f} {\rm e}^{-\tau \widehat{M}_{f}^{2}} \nonumber\\
&&\times\int\frac{d^2 k_{3}}{(2\pi)}\frac{d^{2} k_{b}}{(2\pi)^{2}} {\rm e}^{-\tau ( (\omega^{2}_{n}+ k_{3}^{2}) \frac{{\rm tan}(|Q_{f}E| \tau)}{|Q_{f}E|\tau}+k^{2}_{b}\frac{{\rm tanh}(|Q_{f}B|\tau)}{|Q_{f}B|\tau})},  \label{emt2}
\end{eqnarray}
with $\widehat{M}_f={M}_f (eE, eB,T )$. 
Performing sum over Matsubara frequencies and integrating over $k's$, the gap equation can be written as  
\begin{eqnarray}
\widehat{M}_f &=& m_{f}+ \frac{\alpha_{\rm eff}}{3\pi^{2}}\sum_{f=u,d} \int^{\widehat{\tau}^{2}_{ir}}_{\tau^{2}_{uv}} d\tau \tilde{M}_f {\rm e}^{-\tau \widehat{M}_{f}^{2}}\nonumber\\&& \times\Theta_{3} \bigg(\frac{\pi}{2}, {\rm e}^{- \frac{|Q_{f}E|}{4 T^2{\rm \tan}(|Q_{f}E|\tau)}}\bigg)
\nonumber\\&& \times \frac{|Q_{f}E|}{{\rm tan}(|Q_{f}E|\tau)} \frac{|Q_{f}B|}{{\rm tanh}(|Q_{f}B|\tau)},  \label{emt3}
\end{eqnarray}
where $\Theta_{3}(\frac{\pi}{2},e^{-x})$, is the third Jacobi's theta function.
The quark-antiquark condensate is given by: 
\begin{eqnarray}
-\widehat{\left\langle \psi \bar{\psi} \right\rangle} &=& \frac{3}{4\pi^2}\sum_{f=u,d} \int^{\bar{\tau}^{2}_{ir}}_{\tau^{2}_{uv}} d\tau \widehat{M}_f {\rm e}^{-\tau \widehat{M}_{f}^{2}} \nonumber\\&& \times\Theta_{3} \bigg(\frac{\pi}{2}, {\rm e}^{- \frac{|Q_{f}E|}{4 T^2{\rm \tan}(|Q_{f}E|\tau)}}\bigg)
\nonumber\\&& \times \frac{|Q_{f}E|}{{\rm tan}(|Q_{f}E|\tau)} \frac{|Q_{f}B|}{{\rm tanh}(|Q_{f}B|\tau)}
  \label{emt4}.
\end{eqnarray}
The confinement length scale is now temperature, electric and magnetic field dependent, is of the form:
\begin{eqnarray}
\widehat{\tau}_{ir}=\tau_{ir}\frac{M_f}{\widehat{M}_f}.\label{emt5}
\end{eqnarray} 
First, we consider the case of pure electric field and at finite temperature. The numerical solution of Eq.~(\ref{emt3}) as a function of $T$, for different given values of $eE$, is shown in the Fig.~\ref{Fig10}.
The dynamical mass $\widehat{M}_f$, monotonically decreases with the increase of $T$ until the dynamical chiral symmetry partially restored. The response of $eE$  is to suppress the dynamical chiral symmetry breaking.  The quark-antiquark condensate  Eq.~(\ref{em4}), and the  confinement length scale Eq.~(\ref{em5}), as a function of $T$ for various given values of $eE$, depicted in the Fig.~\ref{Fig11} and Fig.~\ref{Fig12}, respectively.
We see that both the parameters decreases with temperature and at some pseudo-critical temperature $T^{\chi, C}_c$,  the chiral symmetry partially restores, and the deconfinement occurs. We note that the electric field $eE$ suppresses both the parameters not only in the low-temperature region but also reduces the pseudo-critical temperature $T^{\chi, C}_c$.  In  Fig.~\ref{Fig13} and Fig.~\ref{Fig14}, we shows the thermal gradients of the condensate and the confinement length scale respectively, the peaks in the thermal gradients of both parameters shifting towards the lower temperature values upon increasing the value of $eE$. As a result, the critical temperature $T^{\chi, C}_c$ decreases with an increase of electric field strength $eE$, and hence the \textit{inverse electric catalysis} is found at finite temperature in our contact interaction model, which is consistent with the other effective models of QCD \cite{Ruggieri:2016lrn, Wang:2017pje, Tavares:2019mvq}. The magnitude of the critical temperature $T^{\chi,C}_c\approx0.22$, is obtained from the inflection points of the thermal gradients $-\partial_{T}\widehat{\left\langle \psi \bar{\psi} \right\rangle}$, and $\partial _{T} \widehat{\tau}^{-1}_{ir}$.\\
Second, we consider the case of  pure magnetic field strength where now $eE\rightarrow0$), and at finite temperature. We plot the thermal gradient of the quark-aniquark condensate and the confinement length scale respectively, in Fig.~\ref{Fig15} and Fig.~\ref{Fig16}. 
We note that the inflection points in  both the parameters  shifting towards the higher temperatures, it means that magnetic field strength enhances  the pseudo-critical temperature~$T^{\chi,C}_c$, above which the  chiral symmetry partially restored and deconfinement.  Thus, in pure magneitic case and at finite temperature, we find the phenomenon of \textit{magnetic catalysis}, which has been already observed in the contact interaction model~\cite{Ahmad:2016iez,Ahmad:2020jzn}. \\
%For purposes of illustration, we ignore the charge 
%difference of u and d quarks and solve Eq. (6.25) for an
%average common charge q (Klevansky and Leminer,
%1989). 
In  most effective model calculaions of QCD, it is well demonstrated that in order to reproduce the inverse magnetic catalysis effect as predicted by the lattice QCD~\cite{Bali:2012av,Bali:2013esa}, the effective coupling must be taken as magnetic field dependent~\cite{Costa:2015bza,Ruggieri:2016lrn,Ahmad:2016iez,Ahmad:2020jzn} or both temperature and magnetic field dependent~\cite{Farias:2016gmy,Ayala:2015bgv, Ahmad:2020jzn}. In present scenario, we just use the following functional form of the $eB$-dependent effective coupling $\rm{\alpha_{eff}}(eB)$ ~\cite{Ahmad:2016iez}, where the coupling decreases with the magnetic field strength as
\begin{equation}
\rm {\alpha_{eff}}(\kappa)=\rm{\alpha_{eff}} \bigg(\frac{1+a\kappa^{2}+b\kappa^{3}}{1+c\kappa^{2}+d\kappa^{4}}\bigg),\label{emt6}
\end{equation}
here  
$\kappa=eB/\Lambda^2_{QCD}$, with $\Lambda_{QCD}=0.24$ GeV.  The  parameters $a$, $b$, $c$, and  $d$ were extracted  to reproduced the behavior of critical temperature $T^{\chi,C}_{c}$ for the chiral symmetry restoration and deconfinement in the presence of magnetic field strength, obtained by the lattice QCD simulations~\cite{Bali:2012av,Bali:2013esa}.  The thermal gradients of the condensate and the confinement length scale with magnetic field dependent coupling~Eq.~(\ref{emt6})is plotted in the Fig.~\ref{Fig17} and Fig.~\ref{Fig18}, respectively. We see that he critical $T^{\chi,C}_{c}$ , decreases with the increase of $eB$ and thus, we find  the inverse magnetic catalysis at finite $T$. 
\begin{figure}[t!]
\begin{center}
\caption{ %Average Mass; \textit{Mid panel}: 
The behavior of the  $\partial _{T}\widehat{\tau}^{-1}_{ir}$ with magnetic field dependent coupling~Eq.~(\ref{emt6}), as a function of temperature for several values of magnetic field strength $eB$.}
\label{Fig18}
\end{center}
\end{figure}
\\
In the Fig.~\ref{Fig19}, We sketch the combined phase diagram in  the $T^{\chi,C}_{c}-eE,eB$ plane, where we show the inverse electric catalysis, magnetic catalysis and inverse magnetic catalysis (with magnetic dependent coupling). \begin{figure}[t!]
\begin{center}
\includegraphics[width=0.48\textwidth]{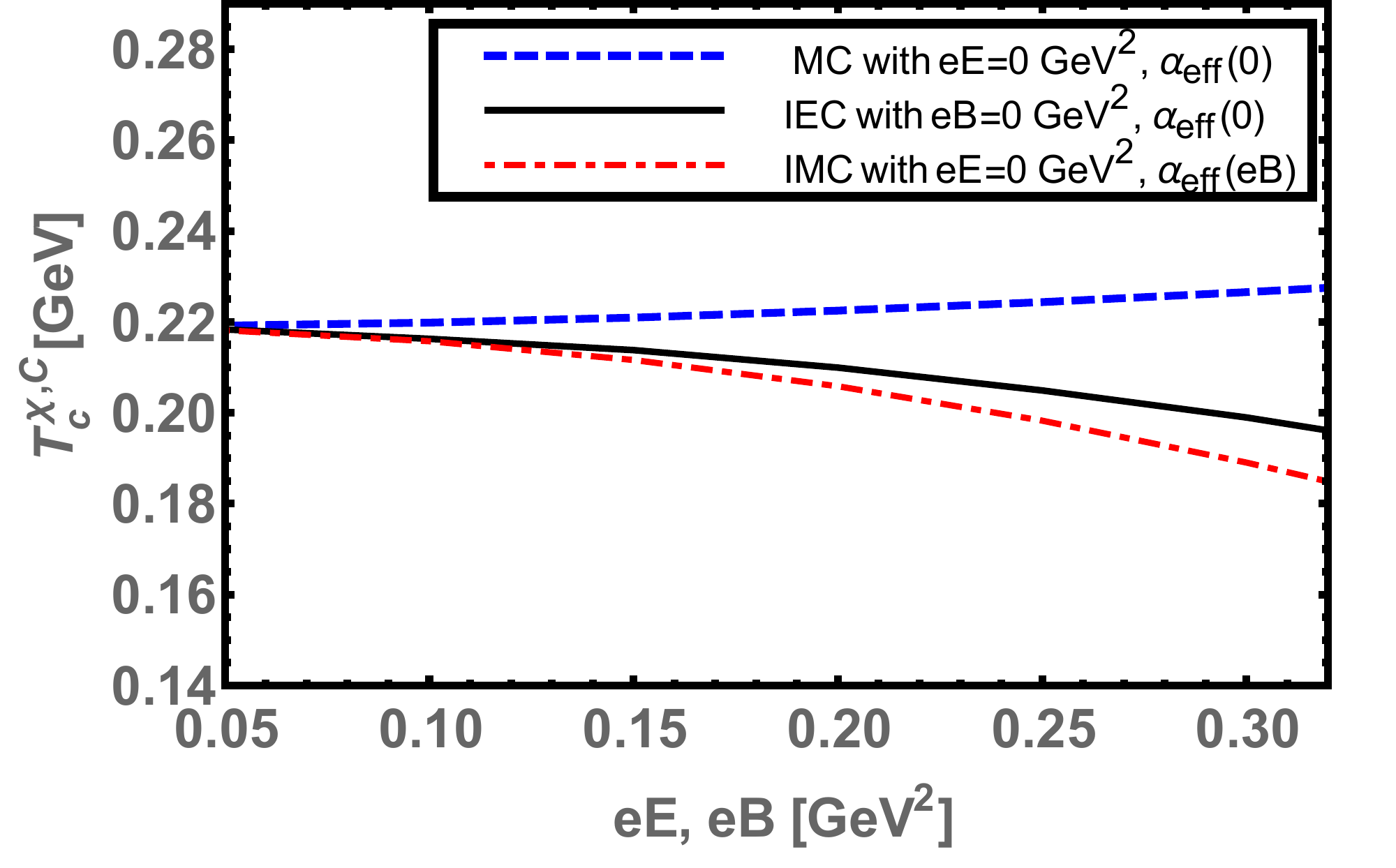}
\caption{ %Average Mass; \textit{Mid panel}: 
The combined phase diagram in the $T^{\chi,C}_{c}$ vs  $eE,eB$  of chiral symmetry breaking and confinement. The plot shows the IEC,  MC and  IMC with $eB$-dependent coupling Eq.~(\ref{emt6}). The  $T^{\chi,C}_{c}$s, are obtained from the inflection points in the $\partial _{T}\widehat{\sigma}$ and $\partial _{T}\widehat{\tau}^{-1}_{ir}$.}
\label{Fig19}
\end{center}
\end{figure} 
In the pure electric background, the solid-black curve represents that the critical temperature $T^{\chi, C}_{c}$ decreases with the increase of $eE$, thus the electric field strength inhibits the chiral symmetry breaking and confinement.  In the pure magnetic limit  (without magnetic field dependent coupling), the magnetic field  $eB$ enhances the critical temperature $T^{\chi, C}_{c}$, and thus $eB$ act as a facilitator of chiral symmetry breaking and confinement.  If we use the magnetic field dependent coupling, we see that the critical temperature $T^{\chi, C}_{c}$  suppresses as the magnetic field strength $eB$ increases (red dotted-dashed curve), now in this case the $eB$ act as an inhibitor of the chiral symmetry and confinement. \\
Third, we take both the non-zero values of  $eE$ and $eB$ and draw the phase diagram in the $T^{\chi,C}_{c}-eE$ plane for various given  values of $eB$ as shown in the Fig.~\ref{Fig20}. Here now the competition between IEC vs MC start: the $eB$ tends to catalyze the chiral symmetry breaking and confinement and as a result $T^{\chi, C}_{c}$ enhances, on the other hand $eE$ try to inhibit the chiral phase transition yield the suppression of $T^{\chi, C}_{c}$, and finally, its ended up with a combined inverse electromagnetic catalysis (IMEC). It may be different for a very strong $eB$, where the $eB$ dominates over the $eE$.   
Next, we consider the  $eB$-dependent coupling Eq.~(\ref{emt6}), and sketch the phase diagram $T^{\chi,C}_{c}$ vs $eE$ for various given values of $eB$. We  see that the effect of parallel $eE$ and $eB$, with $eB$-dependent coupling tends to lower and lower the $T^{\chi,C}_{c}$. Thus, both $eE$ and $eB$ produces IEC, and IMC simultaneously and hence.
\begin{figure}[t!]
\begin{center}
\includegraphics[width=0.48\textwidth]{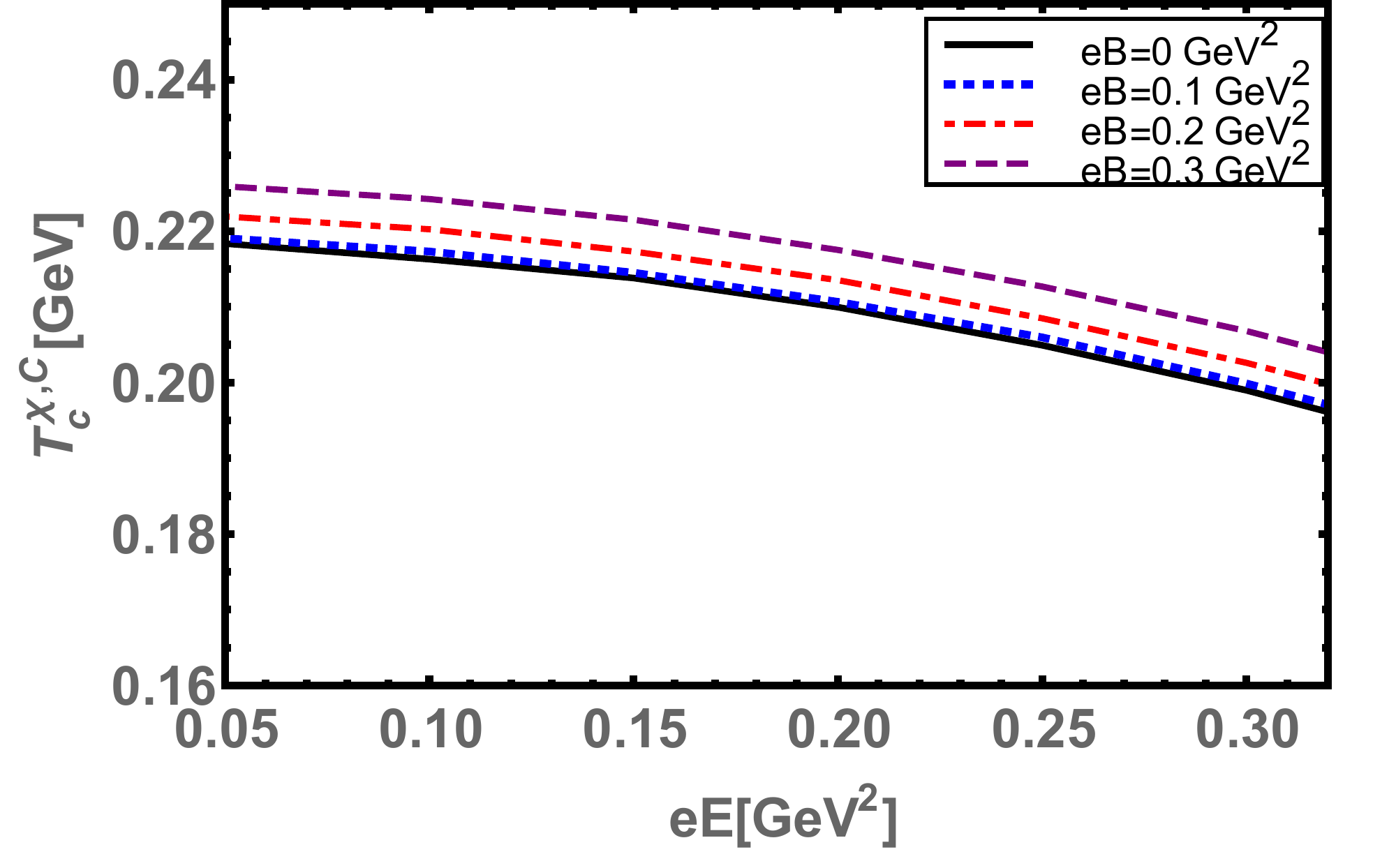}
\caption{ %Average Mass; \textit{Mid panel}: 
The phase diagram in the $T^{\chi,C}_{c}$--$eE$ plane  of chiral symmetry breaking and confinement for various given values of with $eB$. Here $eE$, inhibit the chiral phase transitions, while on the other hand $eB$ facilitates them and as a result we notice the IEMC effect.}
\label{Fig20}
\end{center}
\end{figure}
\begin{figure}[t!]
\begin{center}
\includegraphics[width=0.48\textwidth]{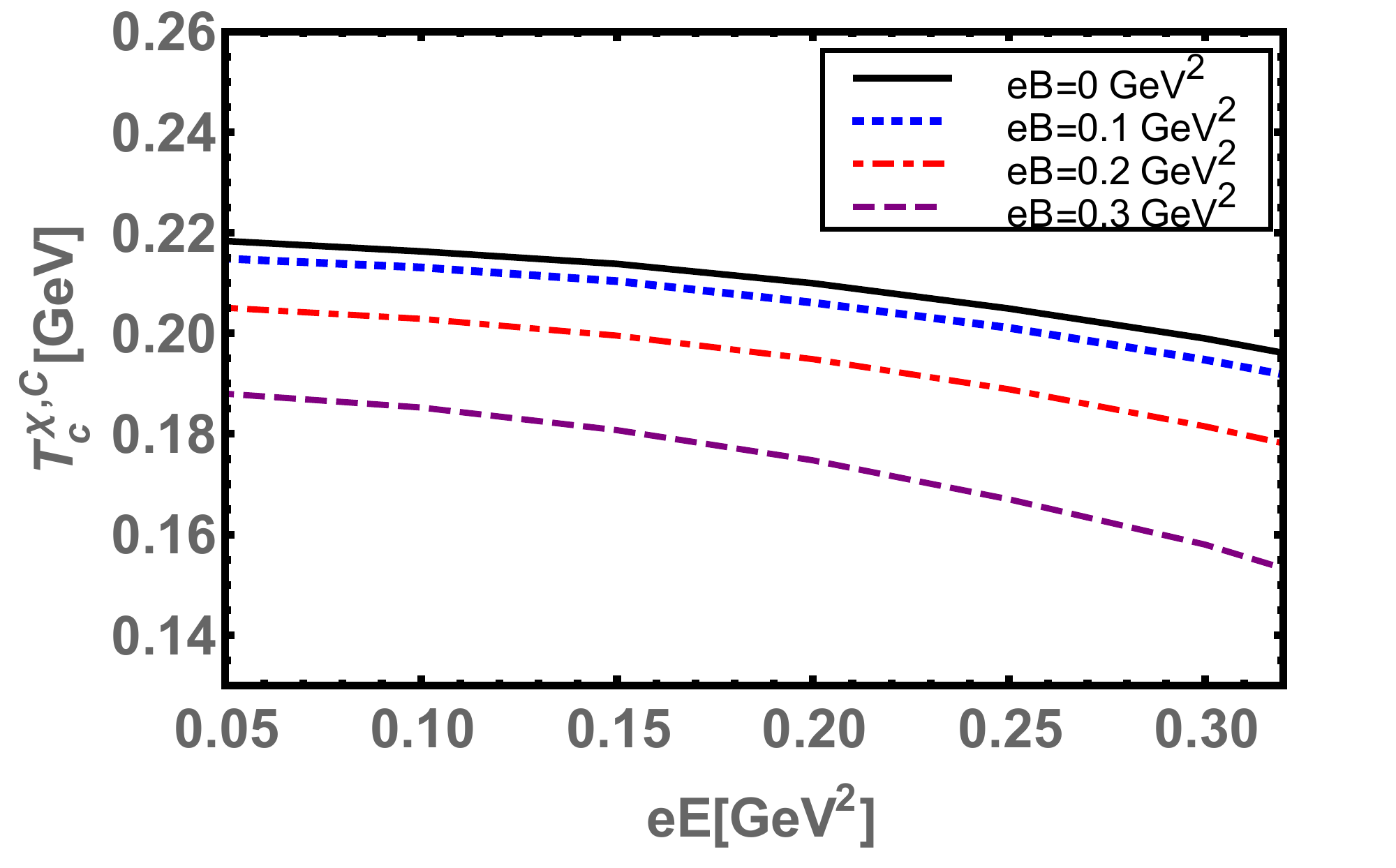}
\caption{ %Average Mass; \textit{Mid panel}: 
The phase diagram in the $T^{\chi,C}_{c}$ vs  $eE$  of chiral symmetry breaking and confinement for various given values of $eE$ with $eB$-dependent couplingEq.~(\ref{emt6}). Here both  fields $eE$ and $eB$ inhibits the chiral phase transitions and hence we noted the IEMC.}
\label{Fig21}
\end{center}
\end{figure}

\section{Summery and Perspectives}\label{conclusions}
In this work, We have studied the influence of uniform, homogeneous, and external parallel electric and magnetic field strength on the chiral phase transitions. In this context, we have implemented the Schwinger-Dyson formulation of QCD, with a gap equation kernel consist of symmetry preserving vector-vector contact interaction model of quarks in rainbow-ladder truncation. We then adopted the well known Schwinger proper-time regularization procedure. 
The outcomes of our work are as follows \\
At zero temperature, in the pure magnetic background, the magnetic field played the role of facilitator of the dynamical chiral symmetry breaking and confinement, and hence, we observed the phenomenon of magnetic catalysis.  On the other hand,  the electric field tends to restore the chiral symmetry and deconfinement above the pseudo-critical electric field $eE^{\chi, C}_c\approx0.34$~GeV$^2$, i.e., the chiral rotation effect revealed in our model. We examined that the electric field acted as an inhibitor of the chiral symmetry breaking and confinement. When  both $eE$  and $eB$, turned on, we find the magnetic catalysis effect for small given values of $eE$ until above the critical electric field strength $eE_{c}\approx0.33$ GeV$^2$, all the order parameters showed  the de Haas-van Alphen oscillatory  type behavior in the region $eB=[0.3,0.54]$ GeV$^2$.   In this particular region, $eE$ dominated over $eB$, and we noticed the electric chiral inhibition effect, while above that region where $eB$ was superior to $eE$, we observed again the magnetic catalysis. We also realized that the pseudo-critical strength $eE^{\chi, C}_{c}$ suppressed with the increase of $eB$, the nature of transitions examined to be of smooth cross-over until above the pseudo-critical magnetic field strength $eB^{\chi, C}_{c}=0.3$ GeV$^2$, the transition suddenly changed to first order. We then located  the position of critical endpoint at $(eB_p=0.3,eE_p=0.33)$ GeV$^{2}$. We further realized that   $eE^{\chi,C}_{c}$ remained constant in the limited region $eB=[0.3-0.54]$ GeV$^2$, and then increased with the larger values of $eB$. \\
In the end, we charted out the phase diagram at finite temperature and in the presence of a parallel electric and magnetic field.  At finite  $T$, in the pure electric limit,  We found that the pseudo-critical temperature decreases as we increase the $eE$, and thus, we perceived the inverse electric catalysis.  On the other hand for pure magnetic field background,  we examined the magnetic catalysis effect in the mean-field approximation and inverse magnetic catalysis with $eB$-dependent coupling. The combined effect of both $eE$ and $eB$ on the  $T^{\chi, C}_c$ yielded the inverse electromagnetic catalysis, with and without $eB-$dependent effective coupling of the model.\\ 
Qualitatively as well as quantitatively, predictions of our CI-model are in agreement with results obtained from the other effective models of QCD and the modern Lattice QCD results. In near future, we plan to extend this work to study the Schwinger pair production rate, the Dynamical chiral symmetry breaking for a higher number of colors, flavors, and in the parallel electromagnetic field. Also, we are interested to study the properties of a light hadron in the background fields. 

\section{Acknowledgments}
I acknowledge A. Bashir and A. Raya for there valuable suggestion and motivation in the process of completion of this work.  I also thanks to the colleagues of the IPE,  Gomal University (Pakistan). 

\bibliographystyle{apsrev4-1} 
\bibliography{aftab}
 
\end{document}